\documentclass{aa}

\usepackage{graphicx}
\usepackage{txfonts}
\usepackage{url}
\usepackage{natbib}
\usepackage{color}
\bibpunct{(}{)}{;}{a}{}{,}
\begin{document}

\title{From grains to pebbles: the influence of size distribution and chemical composition on dust emission properties}

\author{N. Ysard\inst{\ref{inst1}}
\and M. Koehler\inst{\ref{inst1}, \ref{inst2}}
\and I. Jimenez-Serra\inst{\ref{inst3}, \ref{inst2}}
\and A.P. Jones\inst{\ref{inst1}}
\and L. Verstraete\inst{\ref{inst1}}}

\institute{Institut d'Astrophysique Spatiale, CNRS, Univ. Paris-Sud, Universit{\'e} Paris-Saclay, B{\^a}t. 121, 91405 Orsay cedex, France\label{inst1}
\and School of Physics and Astronomy, Queen Mary Univeristy of London,Mile End Road, E1 4NS London\label{inst2}
\and Centro de Astrobiolog{\'i}a (CSIC, INTA), Ctra. de Ajalvir, km. 4, Torrej{\'o}n de Ardoz, 28850 Madrid, Spain\label{inst3}
\\ \email{nathalie.ysard@ias.u-psud.fr}}

\abstract
{The size and chemical composition of interstellar dust grains are critical in setting the dynamical, physical, and chemical evolution of all the media in which they are present. Thanks to facilities such as the Atacama Large Millimeter/submillimeter Array (ALMA) and, in the future, the Square Kilometer Array (SKA), thermal emission in the (sub)millimetre to centimetre domain has become a very convenient way to trace grain properties.}
{Our aim is to understand the influence of the composition and size distribution of dust grains on the shape of their spectral energy distribution (peak position, spectral index) in dense interstellar regions such as molecular clouds, prestellar cores, young stellar objects, and protoplanetary discs.}
{Starting from the optical constants defined in The Heterogeneous dust Evolution Model for Interstellar Solids (THEMIS) for amorphous hydrogenated carbon grains and amorphous silicates in addition to water ice, we defined six material mixtures that we believe are representative of the expected dust composition in dense interstellar regions. The optical properties of 0.01~$\mu$m to 10~cm grains were then calculated with effective medium and Mie theories. The corresponding spectral energy distributions were subsequently calculated for isolated clouds either externally heated by the standard interstellar radiation field alone or in addition to an internal source.}
{The three main outcomes of this study are as follows. Firstly, the dust mass absorption coefficient strongly depends on both grain composition and size distribution potentially leading to errors in dust mass estimates by factors up to $\sim 3$ and 20, respectively. Secondly, it appears almost impossible to retrieve the grain composition from the (sub)millimetre to centimetre thermal emission shape alone as its spectral index for $\lambda \gtrsim 3$~mm does not depend on dust composition. Thirdly, using the `true' dust opacity spectral index to estimate grain sizes may lead to erroneous findings as the observed spectral index can be highly modified by the dust temperature distribution along the line of sight, which depends on the specific heating source and on the geometry of the studied interstellar region.}
{Based on the interpretation of only the spectral shape of (sub)millimetre to centimetre observational data, the determination of the dust masses, compositions, and sizes are highly uncertain.}

\keywords{ISM: general - ISM: dust, extinction - ISM: evolution}
   \authorrunning{}
\titlerunning{From grains to pebbles}
\maketitle
%

\section{Introduction}
\label{introduction}

Interstellar grains constantly evolve throughout the interstellar medium (ISM) in response to their local environment. This evolution occurs from diffuse regions to protoplanetary discs via dense molecular clouds, prestellar cores, and young stellar objects (YSOs). The term `evolution' can be understood in different ways, namely in terms of chemical evolution (e.g. reactions on the grain surface, accretion of gas phase elements), size evolution (e.g. growth by accretion and/or coagulation, decay by erosion, collisional fragmentation, and photoprocessing), or evolution of the grain structure (e.g. compact, porous, single grains, and aggregates). All these facets of grain evolution are important to probe the ISM chemical, physical, and dynamical properties but in the following we focus only on the impact of grain size evolution on the dust spectral energy distribution (SED) at long wavelengths.

Grain growth most probably starts by accretion and/or coagulation in dense molecular clouds. Dust growth is evidenced by a decrease in temperature, an increase in the far-infrared (FIR) and submillimetre opacity, an increase in the selective extinction $R_V = A_V / E(B-V)$, and an increase in the scattering efficiency from the visible to the near-infrared (NIR) \citep[e.g.][]{Whittet2001, Campeggio2007, Pagani2010, Schlafly2016, Remy2017, Remy2018}. Along with these three phenomena, the spectral index of the emissivity in the FIR and submillimetre increases \citep[e.g.][]{Juvela2015}. Two explanations have been put forward to explain this increase: (i) the physics of the emission of amorphous solids corroborated by laboratory experiments on ISM dust grain analogues \citep[e.g. ][]{Agladze1996, Mennella1998, Boudet2005, Meny2007, Coupeaud2011, Demyk2017}; and (ii) a change in the chemical composition of the grain surface \citep{Koehler2015, Jones2016, Ysard2016}. According to all these observational facts and current dust models, grain sizes are not expected to increase much more than by one order of magnitude from $\sim 0.1~\mu$m in the diffuse ISM to $\sim 1~\mu$m in molecular clouds \citep[e.g.][]{Koehler2015, Steinacker2015} along with a structural change from rather compact single grains to fractal or fluffy aggregates.

Denser and more evolved regions exhibit much shallower dust SEDs in the FIR to millimetre wavelength range with spectral indices lower than in molecular clouds and even lower than in the diffuse ISM. This was observed in the first hydrostatic core candidates \citep{Young2018} and YSOs \citep{Chiang2012, Miettinen2012, Miotello2014, Choi2017, Gerin2017, Sheehan2017} with spectral index values below 1.5. Observations of protoplanetary discs also show very low spectral index values \citep[e.g.][]{Isella2010, Kwon2015, Perez2015, Ribas2017, Tripathi2018}, which decrease from the outer to inner disc regions \citep[e.g.][]{Guilloteau2011, Perez2012, Trotta2013, Tazzari2016}. Spectral index values lower than unity have been attributed to the growth of dust particles to radii of around  1mm to 1 cm in size \citep[e.g.][]{Natta2004, Natta2007, Draine2006, Birnstiel2010, Liu2017}. The scheme is that after an initial fractal growth of grains to micronic aggregates, grains continue to grow and are compacted by grain--grain collisions, ram pressure of the gas, and self-gravity for the largest bodies of $\sim 1$~km in size \citep[e.g.][]{Blum2008, Blum2018, Wada2008, Paszun2009, Dominik2016} and drift towards disc centres. The grain chemical composition is also expected to vary inside protoplanetary discs from the mid-plane to the surface and from the inner to outer regions. Most grain growth studies have assumed the same grain composition at all radii and heights \citep[e.g.][]{Perez2012, Tazzari2016}. Therefore the general questions arise of up to what sizes can dust grains grow dependent on local conditions within molecular clouds, prestellar cores, YSOs, and protoplanetary discs, and how the derived sizes  are affected by changes in the dust grain chemical composition. In order to answer these questions, it is necessary to carry out detailed model calculations of dust emission at multiple wavelengths considering realistic dust properties with a wide range of grain compositions, and to compare these theoretical predictions with observations. Instruments such as the Atacama Large Millimeter/submillimeter Array (ALMA) and, in the future, Square Kilometre Array (SKA) will allow us to measure the distribution of centimetre-sized grains down to AU spatial scales, opening the possibility to test grain growth and any dependence on grain composition in dense interstellar regions.

After exploring the effect of complex grain structures with a single chemical composition in \citet{Ysard2018}, the intent of this study is to investigate the effect of complex chemical compositions for isolated spherical grains up to very large sizes. This rather simple approach may not be too unrealistic in the case of very dense environments where grain--grain compaction and gas compression may be efficient processes \citep[e.g.][]{Kataoka2013}. In any case, this study could be used as a basis to compare the influence of composition verses structure complexity on the dust optical property variations.

This paper is organised as follows. Section~\ref{methods} presents the dust chemical compositions considered in this study and the methods used to first translate them into optical properties and then into dust emission spectra. Section~\ref{results} shows the resulting absorption and scattering efficiencies along with the corresponding mass absorption coefficients. Section~\ref{SED} presents the associated emission spectra in the case of optically thin (Sect.~\ref{optically_thin}) and optically thick regions (Sect.~\ref{optically_thick}) in addition to the associated FIR to centimetre spectral indices (Sect.~\ref{ALMA}). Finally, Sect.~\ref{summary} summarises our results.

\section{Dust properties and methods}
\label{methods}

Dust emission depends on grain optical properties and size distribution, and on the strength and intensity of the radiation field. In this section, we describe the variations in dust composition considered in this study and how we translate them into optical properties. The methods used to derive the corresponding dust emission are also outlined.

\subsection{Dust optical properties}
\label{dust_properties}

Our starting point is the optical constants (i.e. complex refractive indices $m = n+{\rm i}k$) from the dust modelling framework THEMIS\footnote{See \url{http://www.ias.u-psud.fr/themis/}.} (The Heterogeneous dust Evolution Model for Interstellar Solids), which is briefly summarised in \citet{Jones2017}\footnote{For the full details of the model see: \citet{Jones2012a, Jones2012b, Jones2012c, Jones2013, Koehler2014, Koehler2015, Jones2016, Ysard2016, Ysard2018}.}. Two main sets of optical constants are included in THEMIS: amorphous magnesium-rich silicates with metallic iron and iron sulphide nano-inclusions and amorphous semi-conducting hydrocarbon grains. For the amorphous silicate component, THEMIS consists of two chemical compositions: one similar to that of enstatite and another similar to that of forsterite \citep{Scott1996}. As the differences are small when moving away from the 10 and 18~$\mu$m silicate spectral features, we only consider forsterite normative composition in the following and refer to it as a-Sil. For the carbonaceous component, we consider two extreme cases in terms of hydrogen content: aromatic-rich grains with an hydrogen fraction $X_{\rm H} \sim 0.02$, referred to as a-C, and aliphatic-rich grains with $X_{\rm H} \sim 0.58$, referred to as a-C:H. In addition to these three materials, we also consider water ice with the optical constants given by \citet{Warren1984}.

Starting from the four aforementioned materials, we consider several composition mixtures and grain structures. For the sake of comparison, we first consider compact grains of purely a-Sil, a-C, or a-C:H. Subsequently, according to \citet{Koehler2015}, we consider compact grains made of two thirds a-Sil and one third a-C (Mix 1) or one third a-C:H (Mix 2), in terms of volume fractions. These allow reproduction of the mass fractions derived by \citet{Jones2013} for the diffuse ISM. The effect of porosity is tested for the Mix 1 mixture, with a porosity degree of 50\% (Mix 1:50). We also evaluate the impact of the presence of a water ice mantle on compact Mix 1 grains (Mix 1:ice). We further consider two material compositions defined in \citet{Pollack1994} based on depletion measurements: (i) 21\% a-Sil and 79\% a-C (Mix 3); and (ii) 8\% a-Sil, 30\% a-C, and 62\% water ice (Mix 3:ice). The various grain compositions are summarised in Table~\ref{table_composition}.
For each grain composition, we derive the absorption and scattering efficiencies $Q_{abs}$ and $Q_{sca}$, respectively, and the asymmetry factor of the phase function $g = \langle {\rm cos}\theta \rangle$. To allow fast calculations, we make the major assumption that the grains are spherical and compute their optical properties using the Mie theory \citep{Mie1908, Bohren1983} with the Fortran 90 version of the BHMIE routine given in \citet{BHMIE}. For grains consisting of two or three materials, we first derive effective optical constants following the Maxwell Garnett mixing rule \citep{MG1904, BHMIE}. Indeed, we assume that in Mix 1 grains, for example, carbon appears as proper inclusions in the silicate matrix rather than assuming a completely random inhomogeneous medium. \citet{Mishchenko2016a, Mishchenko2016b} performed exhaustive studies of the applicability of the Maxwell Garnett mixing rule to heterogeneous particles. These latter authors showed that this rule can provide accurate estimates of the scattering matrix and absorption cross-section of heterogeneous grains at short wavelengths (typically up to the visible for a 0.1~$\mu$m grain and to the mid-infrared (MIR) for a 10~$\mu$m grain) if two criteria are met: both the size parameter of the inclusions and the refractive index contrast between the host material and the inclusions have to be small. Moreover, \citet{Mishchenko2016a} demonstrated that the extinction and asymmetry-parameter errors of the Maxwell Garnett mixing rule are significantly smaller than the scattering-matrix errors, remaining small enough for most typical applications and in particular the kind of applications we perform here. It is however well known that this kind of mixing rule systematically underestimates the absorption efficiency in the FIR to millimetre wavelength range, the implications of which are discussed in Sect.~\ref{section_kappa}. We perform our computations with the {\it emc} routine of V. Ossenkopf\footnote{\url{https://hera.ph1.uni-koeln.de/~ossk/Jena/pubcodes.html}.}. For Mix 1 and Mix 2, we assume a matrix of a-Sil with inclusions of a-C or a-C:H, and for Mix 3 a matrix of a-C with inclusions of a-Sil. For grains surrounded by an ice mantle, the optical properties are derived with the core-mantle Mie theory using the BHCOAT routine given in \citet{BHMIE}.

\begin{table*}[ht]
\centering
\caption{Grain compositions, structures, and volume densities in g/cm$^3$ as described in Sect.~\ref{dust_properties}. The amount of each material composing a grain is given as a fraction of volume. The spectral index column refers to the asymptotic values of the absorption efficiency spectral indices as defined in Sect.~\ref{efficiency}. The references correspond to where the optical constants are defined.}. 
\label{table_composition}
\begin{tabular}{l|ccccccc}
\hline
Name      & Material 1 & Material 2 & Material 3     & Structure   & Density  & Spectral index & References \\
\hline
a-Sil     & a-Sil      &           &                 & compact     & 2.95     & 2.10           & 1 \\
a-C       & a-C        &           &                 & compact     & 1.60     & 1.30           & 2 \\
a-C:H     & a-C:H      &           &                 & compact     & 1.30     & 2.00           & 2 \\
Mix 1     & 2/3 a-Sil  & 1/3 a-C   &                 & compact     & 2.50     & 1.35           & 1,2 \\
Mix 2     & 2/3 a-Sil  & 1/3 a-C:H &                 & compact     & 2.40     & 2.00           & 1,2 \\
Mix 1:50  & 2/3 a-Sil  & 1/3 a-C   &                 & porous 50\% & 1.25     & 1.30           & 1,2 \\
Mix 1:ice & 30\% a-Sil & 15\% a-C  & 55\% ice mantle & compact     & 1.68     & 1.35           & 1,2,3 \\
Mix 3     & 21\% a-Sil & 79\% a-C  &                 & compact     & 1.88     & 1.30           & 1,2,4 \\
Mix 3:ice & 8\% a-Sil  & 30\% a-C  & 62\% ice mantle & compact     & 1.34     & 1.30           & 1,2,3,4 \\
\hline
\end{tabular}
\tablebib{1: \citet{Koehler2014}, 2: \citet{Jones2012a, Jones2012b, Jones2012c}, 3: \citet{Warren1984}, 4: \citet{Pollack1994}.}
\end{table*}

\subsection{Dust size distribution}
\label{size_distribution}

In all our calculations, the grain minimum and maximum sizes are 0.01~$\mu$m and 10~cm. Following \citet{Koehler2015}, we assume a log-normal size distribution:
\begin{equation}
\label{equations_size_distribution}
\frac{dn}{da} \propto \exp\left[ -\frac{1}{2} \left( \frac{\ln(a/a_0)}{\sigma} \right)^2 \right],
\end{equation}
where $a$ is the grain radius. The width $\sigma = 0.7$ is fixed but we allow variations in the centroid $a_0$ from 0.1~$\mu$m to 1~cm to mimic the effect of grain growth. The gas-to-dust mass ratio is fixed\footnote{In a realistic dust model, the gas-to-dust mass ratio should decrease in the case where gas species are accreted onto the grains, for instance in the form of water ice mantles, but this is beyond the scope of this study.} to 100.

The choice of the grain size distribution indeed has a significant impact on both the emission and extinction. However, neither observations nor laboratory experiments have yet made it possible to conclusively determine its shape. Many studies rely on power-law size distributions ($dn/da \propto a^p$), which are consistent with grains resulting from collisional fragmentation cascades \citep[e.g.][]{Davis1990, Tanaka1996}. This is what is typically expected for debris discs for instance. However, recent studies on grain growth during core collapse and in protoplanetary discs have shown that their size distributions are expected to deviate substantially from power laws \citep[e.g.][]{Birnstiel2011, Birnstiel2018}. In particular, based on experimental studies of collisional growth \citep{Weidling2009, Guttler2009, Guttler2010, Windmark2012, Gundlach2015}, the model developed by \citet{Lorek2018} for local growth in protoplanetary discs leads to size distributions close to log-normal functions (see their Figs.~2 and 3). These theoretical and experimental results lead us to choose log-normal distributions rather than the more common power-laws which we do however use for comparison purposes several times in this study.

\subsection{Dust emission}
\label{dust_emission}

Two kinds of calculations are required to arrive at the dust emission spectra, which depend on whether the medium considered is optically thin or thick. In the case of an optically thin medium, the dust emission spectrum is computed with the DustEM\footnote{\url{https://www.ias.u-psud.fr/DUSTEM/}.} code described in \citet{Compiegne2011}. DustEM is a numerical tool, which computes the dust emission and extinction as a function of the grain size distribution and optical properties for a given radiation field.

In the case of an optically thick medium, we use the 3D Monte Carlo radiative transfer model CRT \citep[Continuum Radiative Transfer,][]{Juvela2003, Juvela2005}, which is coupled to DustEM \citep{Ysard2012}. We first assume a filamentary cloud externally heated by the standard interstellar radiation field (ISRF) as defined by \citet{Mathis1983} scaled by the $G_0$ factor. Unless otherwise stated $G_0 = 1$. The cloud is represented by an infinite right circular cylinder with constant density along its axis and the following radial density distribution:
\begin{equation}
\label{equation_rho}
n(r) = \frac{n_C}{1 + (r/H_0)^q},
\end{equation}
where $n_C$ is the central density and $H_0$ the internal flat radius. We define $H_0$ for the cloud to be at equilibrium with its mass per unit length equal to the critical mass defined by \citet{Ostriker1964}: $M_{crit} = 2c_S^2/G$ with $G$ being the gravitational constant. The sound speed, $c_S$, is assumed to be constant and computed for a gas temperature of 12~K. We fix the steepness of the profile to $q = 2$ \citep{Arzoumanian2011}. Second, and in order to assess the effect of internal heating, we consider a spherical cloud with the radial density distribution defined in Eq.~\ref{equation_rho} and a central blackbody radiation source of one solar luminosity and a temperature of 3\,000~K, which is consistent with theoretical predictions of pre-main sequence evolution \citep[see Fig. 3 in][]{Wuchterl2003}. The cloud is also externally illuminated by the ISRF. 

These calculations provide dust temperatures and emission spectra. From the dust emission spectra, we further extract the spectral indices $\beta = \alpha - 2$, where $\alpha$ is the slope of the flux density obtained with linear regressions on the log-log flux density versus frequency plots over 1~GHz frequency steps.

\section{Results: Optical properties}
\label{results}

For the grain chemical compositions detailed in Table~\ref{table_composition}, we now present the corresponding optical properties for sizes between 0.01~$\mu$m and 10~cm as absorption and scattering efficiencies (Sect.~\ref{efficiency}). To illustrate the importance of the size distribution choice, we also show mass absorption coefficients for various log-normal and power-law distributions (Sect.~\ref{section_kappa}).

\subsection{Scattering and absorption efficiencies}
\label{efficiency}

\begin{figure*}[!th]
\centerline{\includegraphics[width=1.3\textwidth]{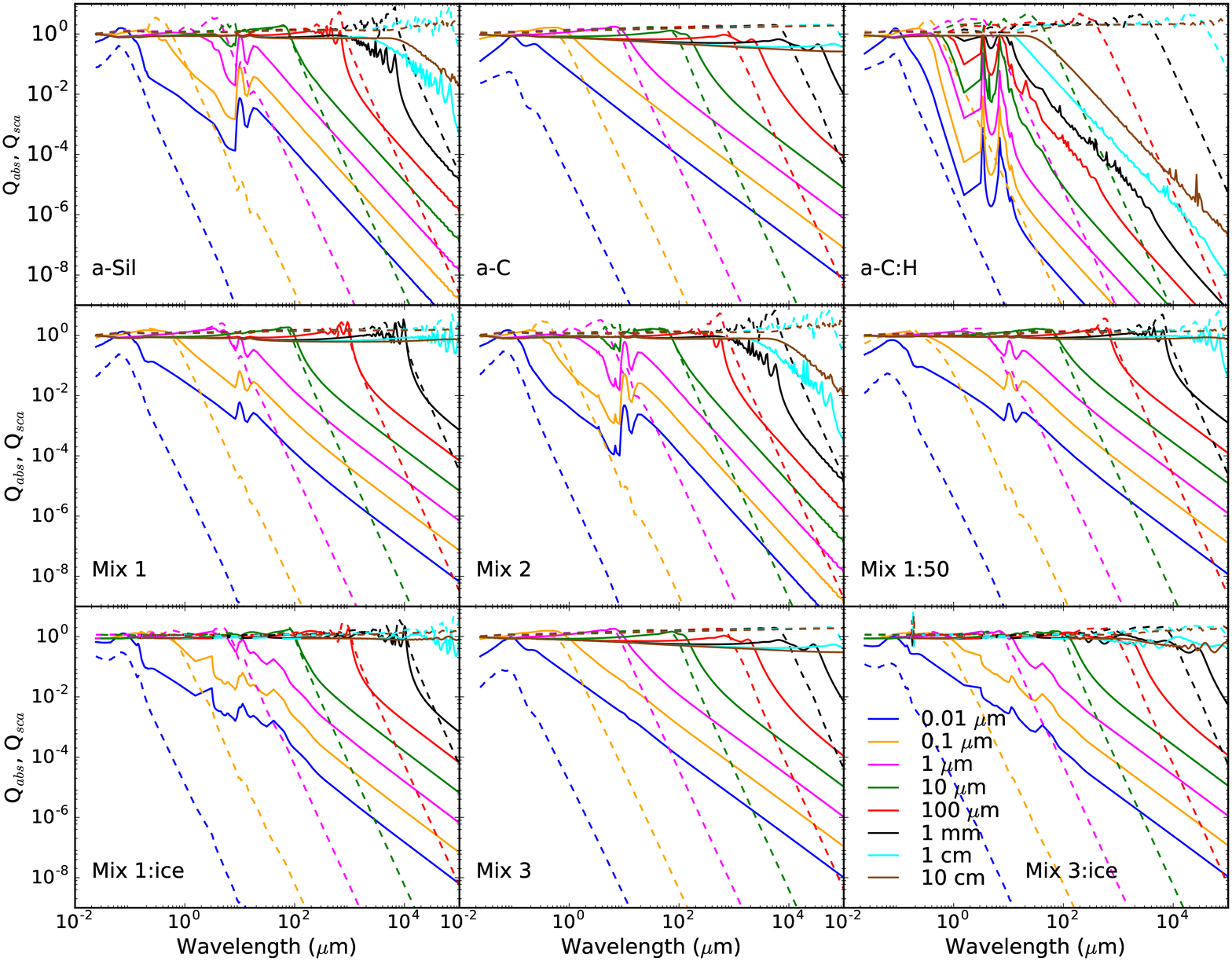}}
\caption{Absorption ($Q_{abs}$, solid lines) and scattering efficiencies ($Q_{sca}$, dashed lines) for the dust compositions described in Sect.~\ref{dust_properties} and Table~\ref{table_composition}. Blue lines show the case of grains with a radius of 0.01~$\mu$m, orange of 0.1~$\mu$m, pink of 1~$\mu$m, green of 10~$\mu$m, red of 100~$\mu$m, black of 1~mm, cyan of 1~cm, and brown of 10~cm.}
\label{figure_efficiencies} 
\end{figure*}

The absorption efficiencies for spherical grains with different radii and material compositions are shown in Fig.~\ref{figure_efficiencies}. For small grains ($a \lesssim 1~\mu$m) consisting of a-Sil and a-C:H, features typical of these materials are visible in the near- to mid-IR. The material mixtures including a-Sil also show the silicate features, except for Mix 3, in which the a-Sil content is too low for the 10 and 18~$\mu$m features to be seen. For grains with ice mantles, the ice spectral features at $\sim 3, 13$, and 50~$\mu$m are visible above the continuum. For larger grains ($1~\mu$m $\lesssim a \lesssim 1~$mm), the spectral features gradually disappear as the continuum to feature contrast increases. Indeed, the {\it threshold wavelength} - under which the absorption and scattering efficiencies tend asymptotically towards one, and above which they sharply decrease - increases with the grain size, its exact spectral position depending on the material conducting properties as discussed in detail in \citet{Ysard2018} for the THEMIS materials.

For grains larger than 1~mm, both the absorption efficiency $Q_{abs}$ and scattering efficiency $Q_{sca}$ are almost constant and equal to one from UV to centimetre wavelengths. This results in relatively flat absorption and scattering efficiencies, only slightly dependent on wavelength. For smaller grains, at wavelengths much larger than the threshold wavelength, the slope of $Q_{abs}$ does not depend on the size but only on the material and is equal to $\sim~$2, 1.3, 2, 1.35, 2, 1.3, 1.35, 1.3, and 1.3 for a-Sil, a-C, a-C:H, Mix 1, Mix 2, Mix 1:50, Mix 1:ice, Mix 3, and Mix 3:ice compositions, respectively. Two interesting points may be noted. Firstly, all grains containing a-C, regardless of the proportion, have almost the same slope as pure a-C grains. Indeed, as a-C grains are low band gap semi-conductors ($E_g = 0.1~$eV for $X_{\rm H} = 0.02$), free-carrier absorption is substantial resulting in longer threshold wavelengths than in the case of a-Sil grains. Secondly, neither porosity nor the presence of ice mantles influence the slope steepness.

\subsection{Mass absorption coefficient}
\label{section_kappa}

\begin{figure*}[!th]
\centerline{\includegraphics[width=1.3\textwidth]{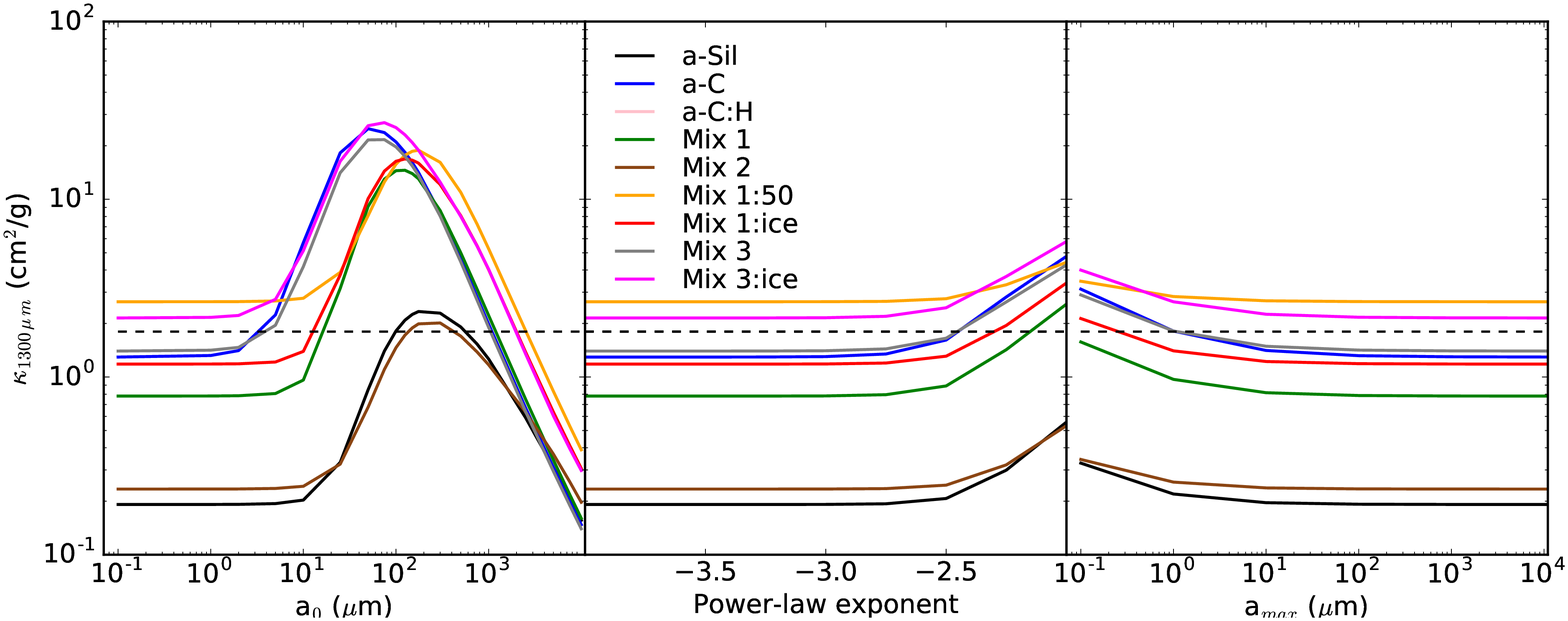}}
\caption{{\it Left:} Mass absorption coefficients $\kappa$ at 1.3~mm for log-normal size distributions centred on $a_0$ and for the dust compositions described in Sect.~\ref{dust_properties} and Table~\ref{table_composition}. Black, blue, pink, green, brown, orange, red, grey, and magenta lines show the cases of a-Sil, a-C, a-C:H, Mix 1, Mix 2, Mix 1:50, Mix 1:ice, Mix 3, and Mix 3:ice grains, respectively. Mass absorption coefficients of pure a-C:H grains are always lower than 10$^{-4}$~cm$^2$/g and thus not visible on the figures. The black dashed line shows $\kappa = 1.8~$cm$^2$/g, a value close to those of the models commonly used in the literature with a gas-to-dust ratio of 100 \citep[e.g.][]{Beckwith1990, Alessio2001, Andrews2009, Birnstiel2018}. {\it Middle:} Same for power-law size distributions $dn/da \propto a^p$ with variable exponent $p$ and $0.01~\mu$m $\leqslant a \leqslant 10$~cm. {\it Right:} Same for power-law size distributions with $p = -3.5$, $a_{min} = 0.01~\mu$m, and variable $a_{max}$.}
\label{figure_kappa} 
\end{figure*}

To infer dust masses from FIR to millimetre observations, observers often use a fixed mass absorption coefficient, $\kappa$, at a given wavelength: see for instance \citet{PLXXII} and \citet{Busquet2019} using the $\kappa$ value of \citet{Beckwith1990} to derive masses of cold molecular cores and protostellar discs, respectively. The left panel of Fig.~\ref{figure_kappa} shows the $\kappa$ values at 1.3~mm for the dust compositions presented in Table~\ref{table_composition} and log-normal size distributions centred at $0.1~\mu$m $\leqslant a_0 \leqslant$ 1~cm (see Eq.~\ref{equations_size_distribution}). As expected, $\kappa$ is highly dependent on grain composition with variations up to a factor of approximately three, with the exception of grains of single composition and those containing a significant volume fraction of a-C:H (i.e. a-Sil, a-C, a-C:H, and Mix 2). However, even greater variations are observed depending on the grain size distribution for $a_0 \geqslant 5-10~\mu$m ($\kappa$ is almost constant otherwise). Excluding grains of single composition or grains containing a-C:H, for $a_0 = 1$ to $\sim 100~\mu$m, $\kappa$ increases by factors of $\sim 7$ to 19. For size distributions centred at larger sizes, $\kappa$ decreases again to $\kappa[a_0 = 1$~cm$] \sim 0.1-0.25 \times \kappa[a_0 = 0.1~\mu$m$]$. This is simply explained by the fact that log-normal size distributions have a characteristic grain size dominating the total dust mass and by the definition of the mass absorption coefficient which, at zero order, can be approximated by $\kappa = (3/4\rho) \times (Q_{abs}/a_0)$. For $a_0 \lesssim 5$ to $10~\mu$m depending on the grain composition, $a_0$ is small compared to the wavelength and the increase in $Q_{abs}(\lambda \gg a_0)$ is proportional to $a_0$ resulting in a constant $\kappa$. For $5-10 \lesssim a_0 \lesssim 60-200~\mu$m, the grain sizes become comparable to the wavelength, resulting in a stronger increase in $Q_{abs}$ and therefore in an increase in $\kappa$. For larger grain sizes, at $\lambda = 1.3~$mm, $Q_{abs} \sim 1$ is almost independent of size for all dust compositions (see Fig.~\ref{figure_efficiencies}) resulting in a decreasing $\kappa$ for increasing sizes ($\kappa \propto 1/a_0$).

For the sake of comparison, the middle and right panels of Fig.~\ref{figure_kappa} again show the $\kappa$ values at 1.3~mm but in the case of power-law size distributions: $dn/da \propto a^p$. The middle panel shows the case of a variable power-law exponent $-4 \leqslant p \leqslant -2$ for constant size limits $0.01~\mu{\rm m} \leqslant a \leqslant 10~$cm, whereas the right panel shows the case of a constant exponent $p = -3.5$ with an increasing maximum grain size $0.01~\mu{\rm m} \leqslant a \leqslant a_{max} = 0.1~\mu$m to 10~cm. Indeed, for the most part, previous studies have used such size distributions, usually with $p = -3.5$ \citep[e.g.][]{Natta2004, Draine2006, Isella2012, Liu2017}. For $p < -2.25$ or $a_{max} \geqslant 1~\mu$m and whatever the grain composition, the mass absorption coefficient is constant and roughly equal to the mass absorption coefficients found for log-normal size distributions centred on $a_0 \leqslant 5~\mu$m. An increase in $\kappa$ of about a factor of 1.3 to 3 is found only when $p \geqslant -2.5$ or $a_{max} < 1~\mu$m, far from the highest values found for the log-normal case at $a_0 \sim 60 - 100~\mu$m. Tables giving the mass absorption coefficient values $\kappa$ for wavelengths from 250~$\mu$m to 2.8~mm are available in Appendix~\ref{appendix_kappa}.

As already stated in the introduction, several laboratory experiments on ISM grain analogues have also shown that their mass absorption coefficient depends not only on their size and composition but also on their temperature \citep[e.g. ][]{Agladze1996, Mennella1998, Boudet2005, Meny2007, Coupeaud2011, Demyk2017}. For instance, \citet{Demyk2017} found that for Mg-rich amorphous silicate analogues, $\kappa_{\rm 1mm}$ doubles when the temperature increases from 10 to 300~K, while the spectral index at submillimetre wavelengths decreases from about 2.2 to 1.6. Even if  the inclusion of such effects is beyond the scope of our study, it should be kept in mind that when observing regions encompassing large temperature gradients, intrinsic variations in the mass absorption coefficient of a given material can introduce significant uncertainties on mass determinations based solely on the long-wavelength FIR to millimetre dust emission.

Further, when calculating the grain optical properties, we make the major assumption that these grains are spherical. However, it is well known that their exact shape significantly alters the scattering and absorption efficiencies and therefore their mass absorption coefficient at all wavelengths. This was shown in studies based on interstellar dust, cometary dust, and terrestrial aerosols. This alteration depends in particular on the size and shape of the monomers composing the aggregate grain \citep[e.g.][]{Liu2015, Wu2016, Min2016, Ysard2018}, on the monomer composition and composition mixing \citep[e.g.][]{Koehler2012, Min2016}, and on the contact area between the monomers \citep{Koehler2011}. All these studies have focused on different points, but compared to the simple case of the porous or non-porous sphere, a common observation can be made: the influence of the exact shape of the aggregates can increase the FIR to millimetre mass absorption coefficient by a few tens of percent and give a lower spectral index in the same wavelength range\footnote{For more details about the comparison of exact computation methods with common approximate methods, we refer the reader to \citet{Min2016} and also to the exhaustive study of \citet{Tazaki2018}.}.

To conclude, Fig.~\ref{figure_kappa} and previous laboratory and theoretical studies clearly illustrate how the choice of a mass absorption coefficient $\kappa$ to infer dust masses from observations is very model-dependent. Thus, inter-comparisons between the dust masses derived from different studies should always be taken with caution and, in particular, it should be noted that the need to adopt a dust model automatically introduces at least one order of magnitude uncertainty into any dust mass determination.

\section{Results: Emission spectra}
\label{SED}

The purpose of this section is to provide numerical values comparable to those derived from astronomical observations. From the optical properties and size distributions presented in the previous section, we therefore calculate SEDs for different radiation fields and their associated spectral indices. 

\subsection{Optically thin medium}
\label{optically_thin}

\begin{figure}[!t]
\centerline{\includegraphics[width=0.45\textwidth]{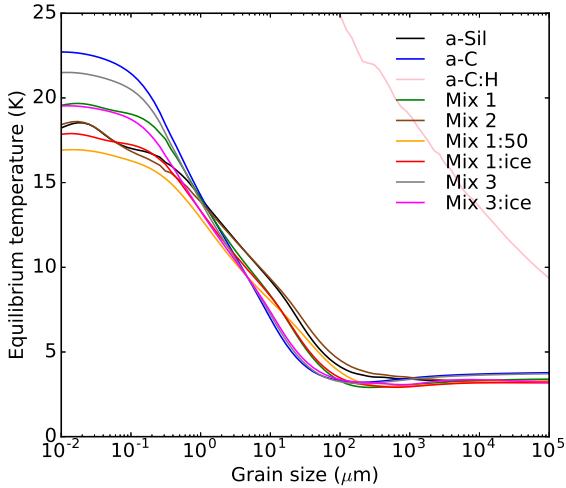}}
\caption{Equilibrium temperature as a function of grain size for the dust compositions described in Sect.~\ref{dust_properties} and Table~\ref{table_composition}, illuminated by the ISRF with $G_0 = 1$. The  same colour code is used as in Fig.~\ref{figure_kappa}.}
\label{figure_ISRF_temperatures} 
\end{figure}

\begin{figure*}[!th]
\centerline{\includegraphics[width=1.3\textwidth]{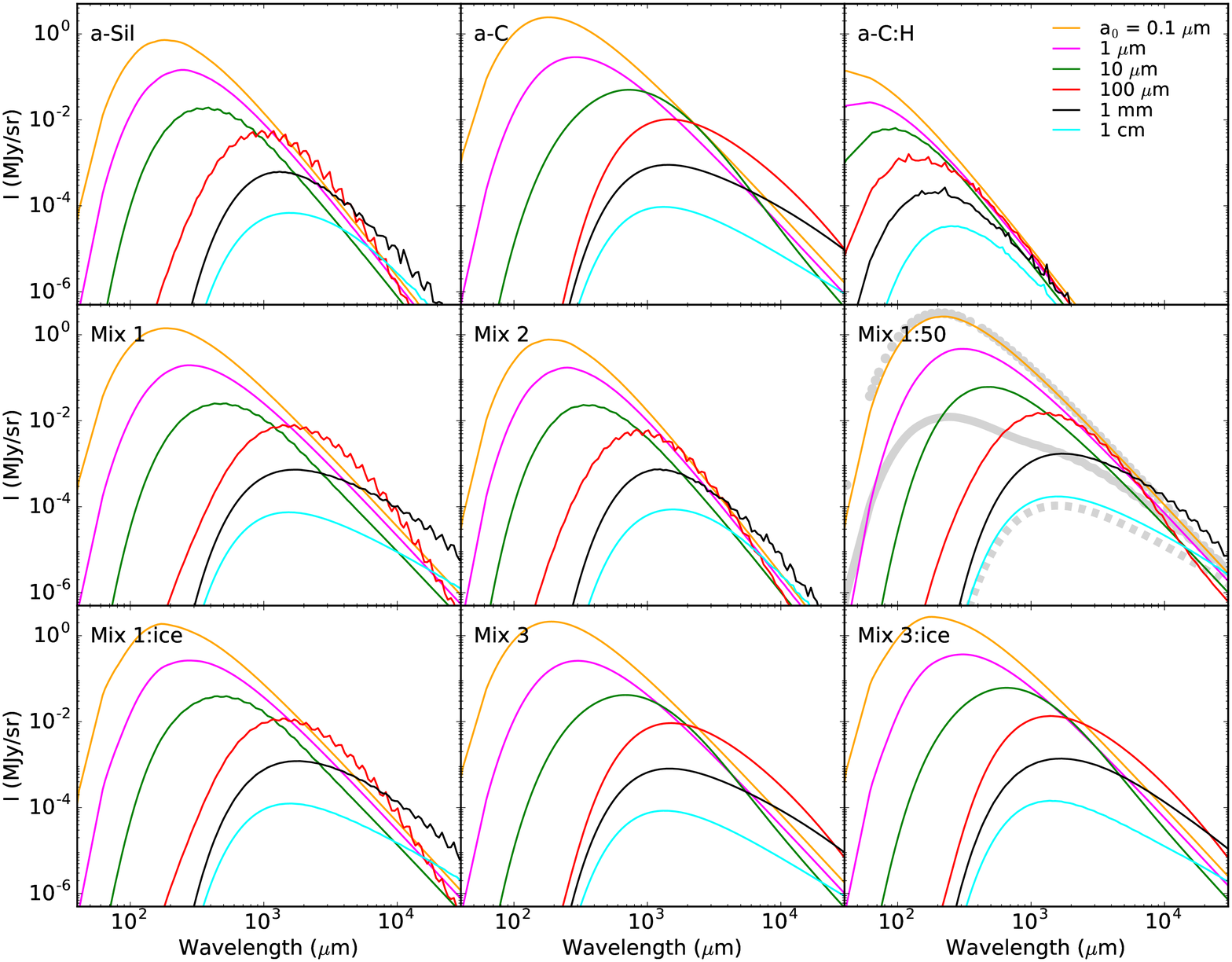}}
\caption{Spectral energy distributions for the dust compositions described in Sect.~\ref{dust_properties} and Table~\ref{table_composition}. The SEDs were obtained considering an optically thin medium with a column density $N_{\rm H} = 10^{20}$~H/cm$^2$ illuminated by the standard ISRF (see Sect.~\ref{dust_emission} for details). The orange lines show the case where the centroid of the log-normal size distribution is $a_0 = 0.1~\mu$m, the pink lines where $a_0 = 1~\mu$m, the green lines where $a_0 = 10~\mu$m, the red lines where $a_0 = 100~\mu$m, the black lines where $a_0 = 1$~mm, and the cyan lines where $a_0 = 1$~cm. The thick grey curves on the Mix 1:50 plot present the case of power-law size distributions with $a_{min} = 0.01~\mu$m and where the dotted line shows the case of $p = -3.5$ and $a_{max} = 1~\mu$m, the solid line of $p = -3.5$ and $a_{max} = 10~$cm, and the dashed line of $p = -2$ and $a_{max} = 10~$cm.}
\label{figure_ISRF_SEDs} 
\end{figure*}

Using the DustEM code, the equilibrium temperatures of the grains presented in Table~\ref{table_composition} and their SEDs as flux densities have been calculated and are shown in Figs.~\ref{figure_ISRF_temperatures} and \ref{figure_ISRF_SEDs}, respectively. As described in Sect.~\ref{methods}, the temperatures and spectra are computed for the standard ISRF with $G_0 = 1$ for various values of the size distribution centroid: $a_0 = 0.1~\mu$m, 1~$\mu$m, 10~$\mu$m, 100~$\mu$m, 1~mm, and 1~cm.

\begin{figure}[!t]
\centerline{\includegraphics[width=0.45\textwidth]{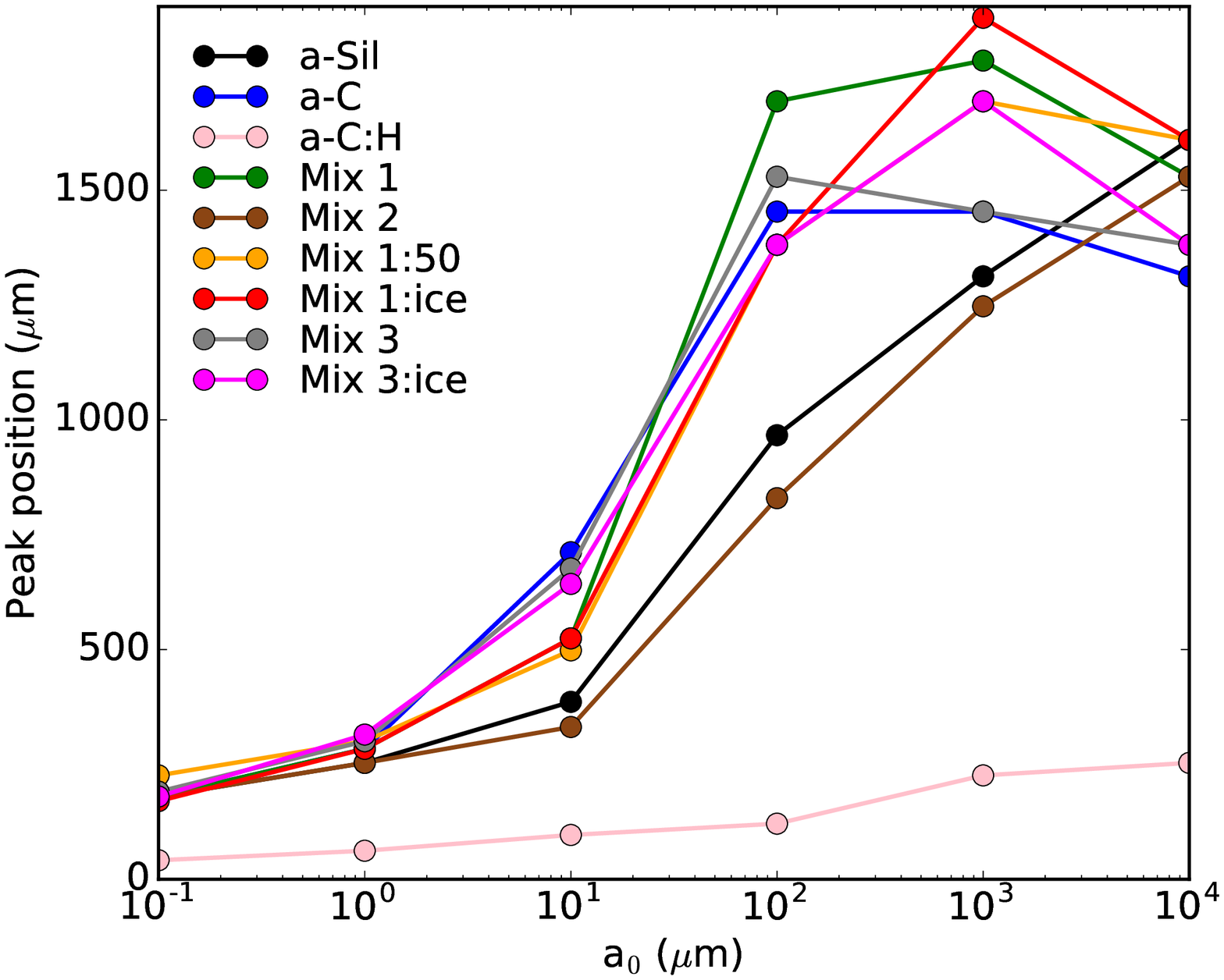}}
\caption{Wavelength of the maxima of the SEDs shown in Fig.~\ref{figure_ISRF_SEDs} as a function of the size distribution centroid $a_0$. The  same colour code is used as
in Fig.~\ref{figure_kappa}.}
\label{figure_ISRF_SEDs_peak_positions} 
\end{figure}

First of all, pure a-C:H grains have a far different behaviour from the other grain types. Indeed, Fig.~\ref{figure_efficiencies} shows that their optical properties are quite different with an absorption efficiency always inferior to the scattering efficiency, regardless of the grain size. This makes these grains poorly emissive which results in high grain equilibrium temperatures for all sizes (Fig.~\ref{figure_ISRF_temperatures}) and flux densities peaking at relatively short wavelengths (Figs.~\ref{figure_ISRF_SEDs} and \ref{figure_ISRF_SEDs_peak_positions}). Although this behaviour is interesting from a theoretical point of view, we do not discuss this case further since such grains are improbable in the ISM because we expect mixing with silicate or ice components and/or partial dehydrogenation due to UV photons\footnote{The optical properties of pure a-C:H grains are discussed in details in \citet{Jones2016} and \citet{Ysard2018}.}.

Figure \ref{figure_ISRF_temperatures} shows the equilibrium temperature as a function of grain size for the grain compositions described in Table~\ref{table_composition}. At small sizes, the temperature depends on the material through its heat capacity and the size-dependent variations in $Q_{abs}$. On the other hand, at larger sizes ($a \gtrsim 100~\mu$m), the temperature is practically independent of the material, which is explained by the fact that all materials, except pure a-C:H, have $Q_{abs} \sim Q_{sca} \sim 1$. Excluding pure a-C:H, the temperature of the largest grains does not vary by more than 1~K from one material to another. A last interesting point regarding the equilibrium temperature is the fact that for pure a-C grains and grains made partly of a-C, the equilibrium temperature slightly increases by about 0.6~K from $a = 100~\mu$m to 1~cm. This is the consequence of the spectral shape of the a-C $Q_{abs}$, which instead of being almost constant for the largest sizes, decreases slightly up to the threshold wavelength, which makes the largest grains slightly less emissive and thus marginally hotter.

\begin{figure*}[!th]
\centerline{\includegraphics[width=1.3\textwidth]{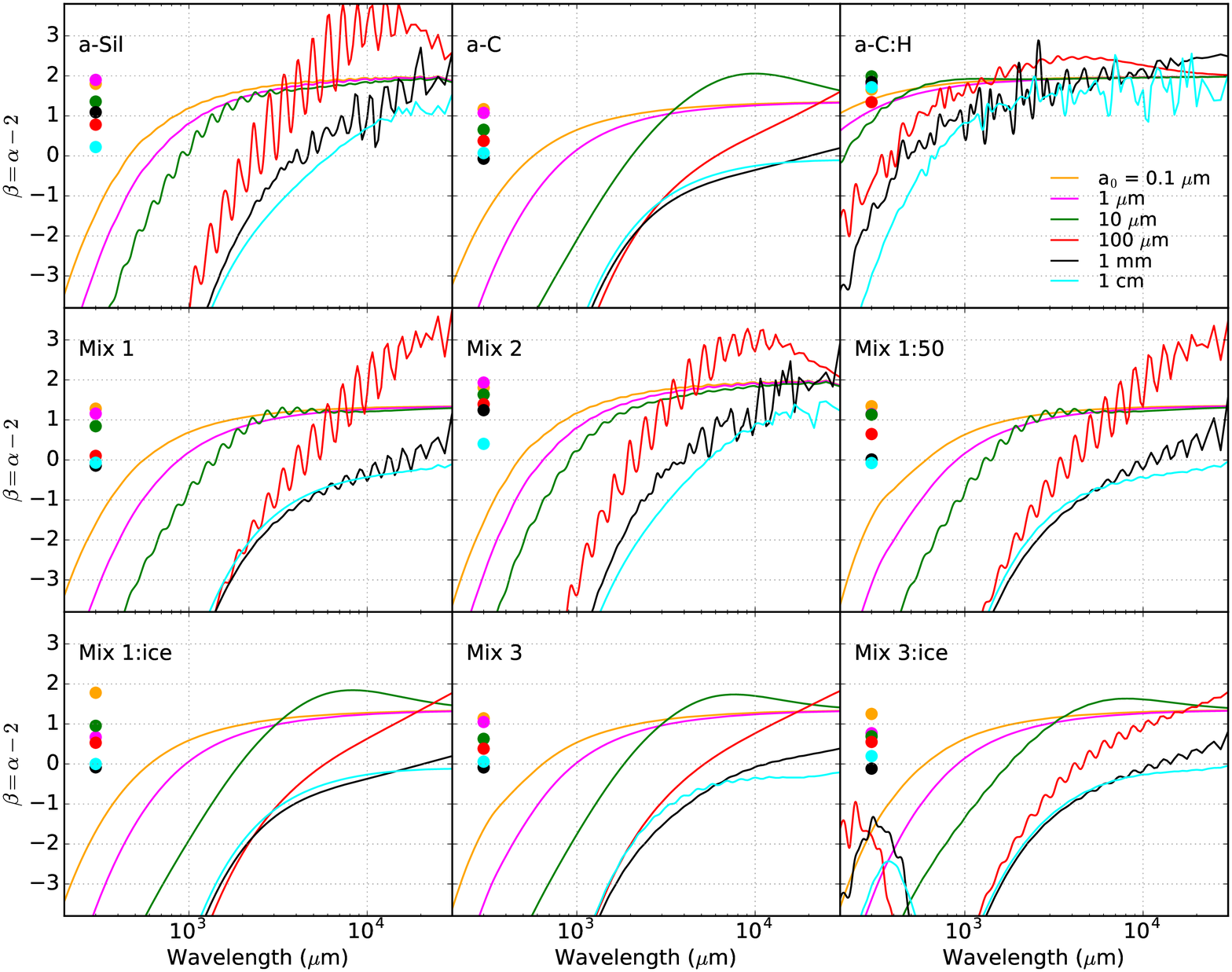}}
\caption{Spectral indices of the SEDs presented in Fig.~\ref{figure_ISRF_SEDs} (optically thin medium), calculated over 1~GHz frequency steps (solid lines). The circles display the spectral indices calculated using a modified blackbody in the Wien part of the SEDs. The  same colour code is used as
in Fig. \ref{figure_ISRF_SEDs}.}
\label{figure_ISRF_SEDs_slopes} 
\end{figure*}

The dependence of temperature on size and material is reflected in the flux density. For grains without a-C (a-Sil, a-C:H, and Mix 2), the wavelength of the SED peak increases with the centroid size $a_0$, whereas for grains containing a-C (i.e. all except a-Sil, a-C:H, and Mix 2), it first increases before decreasing slightly or stagnating when $a_0 \gtrsim 100~\mu$m or 1~mm depending on the grain composition (Fig.~\ref{figure_ISRF_SEDs_peak_positions}). Figure~\ref{figure_ISRF_SEDs} also shows that the slope of the flux density in the FIR and submillimetre strongly depends on both the grain size and composition with steeper slopes for the smallest $a_0$ values. This is  easily visible in Fig.~\ref{figure_ISRF_SEDs_slopes} where $\beta = \alpha - 2$ is plotted for $\lambda \geqslant 200~\mu$m. For the three smallest centroid values, $a_0 = 0.1$, 1, and 10~$\mu$m, $\beta(\lambda \gtrsim 1~{\rm cm})$ is asymptotically equal to the slope of the material absorption efficiency $Q_{abs}$ (see Table~\ref{table_composition}). For size distributions centred on sizes $a_0$ larger than 10~$\mu$m, the spectral indices are globally lower as long as $\lambda \lesssim 10 a_0$. The break in the dependence of the equilibrium temperature on the grain size for $a \sim 100~\mu$m (see Fig.~\ref{figure_ISRF_temperatures}) directly impacts the spectral index. Compared to the smaller ones, the largest grains have slightly higher temperatures when they contain a-C or are of a comparable temperature otherwise. Their contribution is enhanced when $a_0$ increases which distorts the spectral shape of the flux density peak compared to smaller $a_0$ cases. This results in low $\beta$ values increasing with wavelength. Based solely on (sub)millimetre/centimetre $\beta$ measurements, for $a_0$ smaller than 100~$\mu$m, it would be very difficult to determine the grain sizes for most of the grain compositions considered in this study. In Sect.~\ref{optically_thick}, we investigate to what extent this statement holds in optically thick media in which dust growth is expected to occur. 

In addition, it should be noted that the negative spectral index values at short wavelengths are artificial and result from the application of the Rayleigh-Jeans approximation outside its validity domain, that is, too far ahead or too close to the SED peak (see Fig.~\ref{figure_ISRF_SEDs_peak_positions}). In order to fully characterise the SEDs, Fig.~\ref{figure_ISRF_SEDs_slopes} also shows the spectral indices obtained by fitting the Wien part of the SEDs with a modified blackbody: $I_{\nu} \propto B_{\nu}(T) \nu^{\beta}$, where $B_{\nu}(T)$ is the Planck function at temperature $T$ and $\beta$ is the SED spectral index. The wavelength range corresponding to the Wien part of the SEDs depends on both the dust composition and size distribution and is therefore different for all the cases presented in Fig.~\ref{figure_ISRF_SEDs_slopes}. Our choice is to perform the fits on wavelength ranges matching the full width at half maximum of each SED. Positive spectral indices are found for the smallest sizes and null for the largest ($a_0 = 1$~mm and 1~cm). It also appears that for $a_0 \leqslant 10~\mu$m the spectral indices of the modified blackbodies are close to the material absorption efficiencies $Q_{abs}$ slopes (see Table~\ref{table_composition}).

For the sake of comparison, Fig.~\ref{figure_ISRF_SEDs} shows the three extreme cases of power-law size distributions already presented in Sect.~\ref{section_kappa}: (i) $p = -3.5$, $0.01~\mu$m $\leqslant a \leqslant 10$~cm (solid) ; (ii) $p = -2$, $0.01~\mu$m $\leqslant a \leqslant 10$~cm (dashed) ; (iii) $p = -3.5$, $0.01~\mu$m $\leqslant a \leqslant 1~\mu$m (dotted). This comparison leads to two main results. Firstly, case (ii), in which a large part of the mass is distributed in the largest grains, gives a SED similar to the case where $a_0 = 1$~cm. Secondly, case (iii), which places most of the mass in the smallest grains, is similar to the case where $a_0 = 0.1~\mu$m. Both the peak of the SEDs and their long wavelength spectral indices are similar. Case (i) is intermediate where no size dominates the grain size distribution. This results in a very broad SED peaking at about the same wavelength as when $a_0 = 0.1~\mu$m but its centimetre emission in terms of spectral index is close to that of the case where $a_0 = 1$~mm.

\begin{figure*}[!t]
\centerline{\begin{tabular}{cc}
\includegraphics[width=1\textwidth]{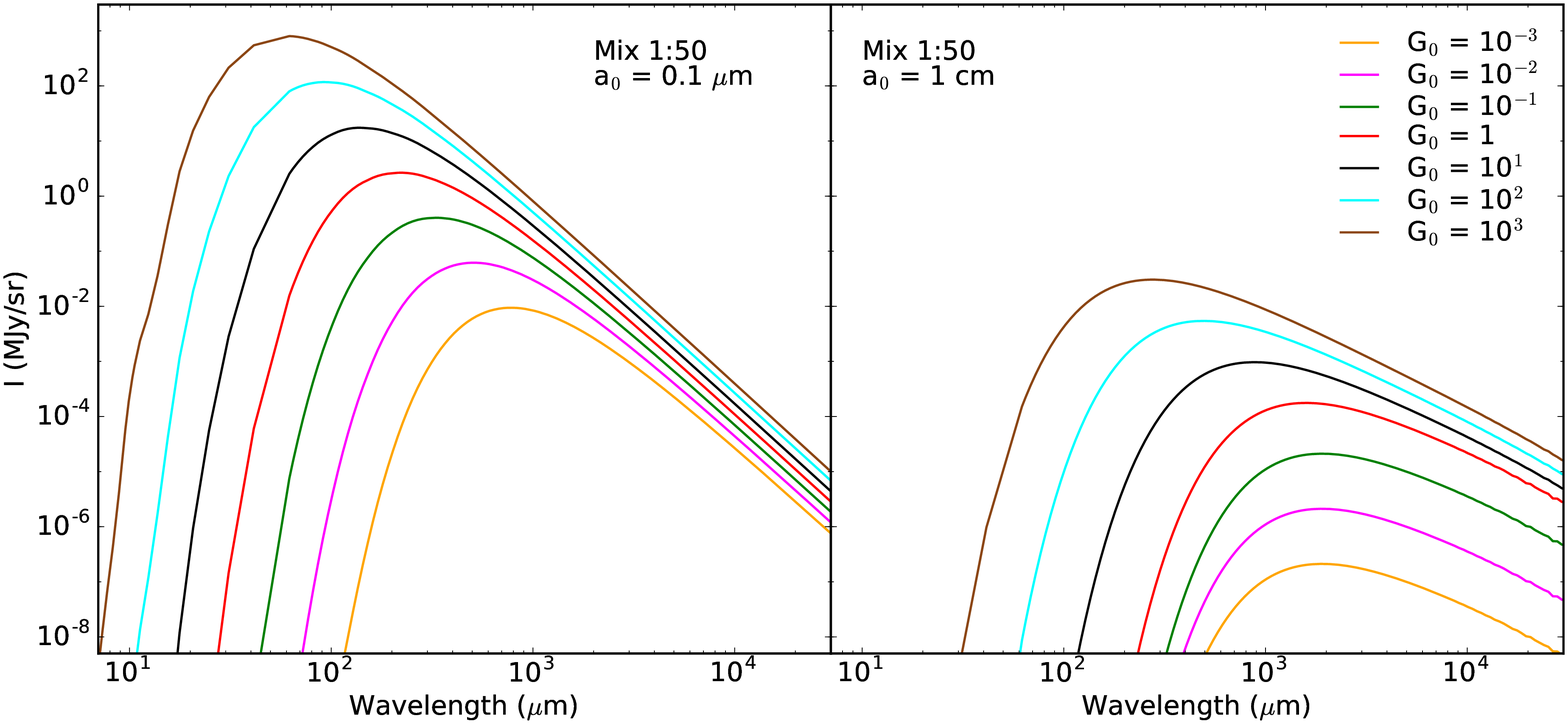} \\
\includegraphics[width=1\textwidth]{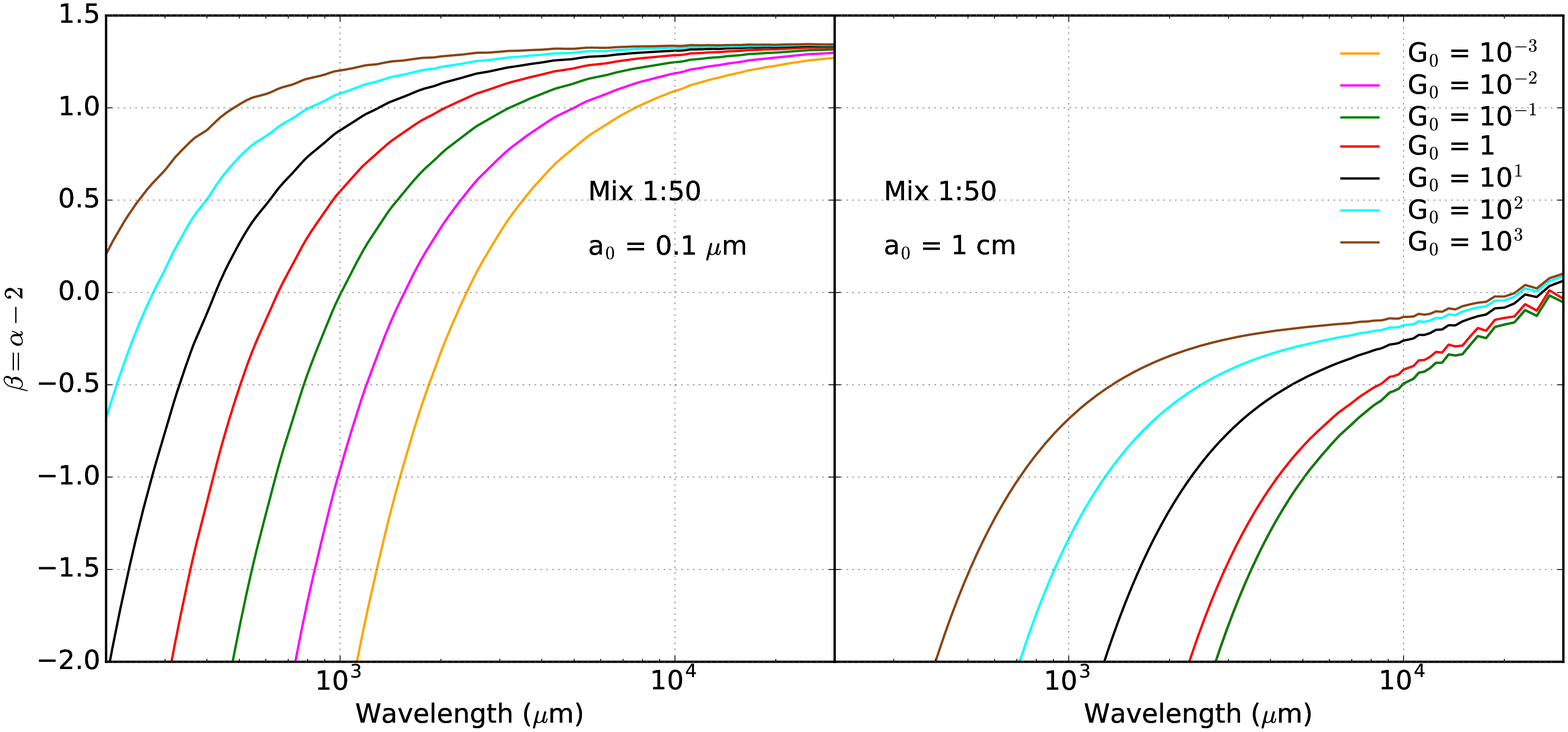}
\end{tabular}}
\caption{{\it Top:} SEDs for the dust composition Mix 1:50 described in Sect.~\ref{dust_properties} and Table~\ref{table_composition}. The SEDs were obtained considering an optically thin medium with a column density $N_{\rm H} = 10^{20}$~H/cm$^2$ illuminated by the standard ISRF scaled by $G_0 = 10^{-3}, 10^{-2}, 10^{-1}, 1, 10, 10^2$, and $10^3$ plotted in orange, magenta, green, red, black, cyan, and brown, respectively. {\it Bottom:} Corresponding spectral indices calculated over 1~GHz frequency steps. The spectral indices for $a_0 = 1$~cm and $G_0 = 10^{-3}, 10^{-2}$, and $10^{-1}$ are overlapping.}
\label{figure_ISRF_SEDs_Go} 
\end{figure*}

Figure~\ref{figure_ISRF_SEDs_Go} shows the influence of the intensity of the radiation field on the SED for $a_0 = 0.1~\mu$m and 1~cm. For $a_0 = 0.1~\mu$m, the peak of the SED shifts to longer wavelengths when $G_0$ decreases. Thus, the spectral index reaches its asymptotic value, equal to the slope of the material absorption efficiency $Q_{abs}$, at shorter wavelengths for higher $G_0$. For $a_0 = 1$~cm, the peak of the SED also shifts to longer wavelengths when $G_0$ decreases provided that $G_0 \geqslant 0.1$. For lower $G_0$, there are no longer enough energetic photons available to warm up the very large grains which make most of the mass in the size distribution and these grains end up being at equilibrium with the cosmic microwave background (CMB) temperature. This break starts appearing for $a_0 \gtrsim 1$~mm and $G_0 \lesssim 10^{-2}$.

\subsection{Optically thick medium}
\label{optically_thick}

\begin{figure*}[!th]
\centerline{\includegraphics[width=1.3\textwidth]{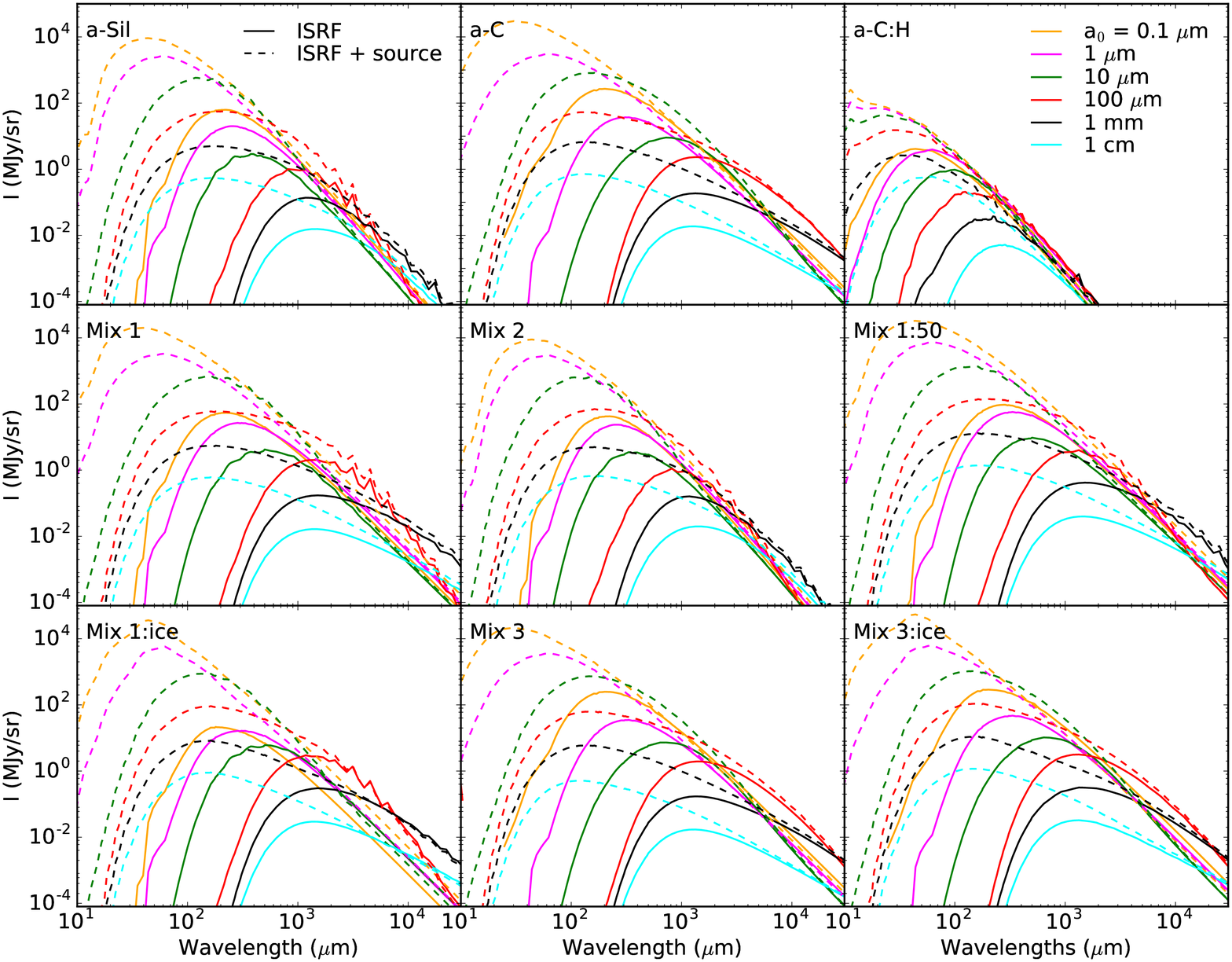}}
\caption{Spectral energy distributions for the dust compositions described in Sect.~\ref{dust_properties} and Table~\ref{table_composition}. The SEDs were obtained considering a cloud with central density $n_C = 10^5$~H/cm$^3$ (Eq.~\ref{equation_rho}), illuminated either by the standard ISRF alone (solid lines) or in addition to an internal heating source (dashed lines, see Sect.~\ref{dust_emission} for details). The orange lines show the case where the centroid of the log-normal size distribution is $a_0 = 0.1~\mu$m, the pink lines where $a_0 = 1~\mu$m, the green lines where $a_0 = 10~\mu$m, the red lines where $a_0 = 100~\mu$m, the black lines where $a_0 = 1$~mm, and the cyan lines where $a_0 = 1$~cm.}
\label{figure_CRT_SEDs} 
\end{figure*}

As described in Sect.~\ref{dust_emission}, we compute the emission spectra at the centre of an optically thick cloud (see Eq.~\ref{equation_rho}), either externally illuminated by the ISRF alone or with an additional central heating source. The resulting SEDs for various size distribution centroids 0.1~$\mu$m $\leqslant a_0 \leqslant 1$~cm and $n_C = 10^5$~H/cm$^3$ are presented in Fig.~\ref{figure_CRT_SEDs}. 

\begin{figure*}[!th]
\centerline{\includegraphics[width=1.3\textwidth]{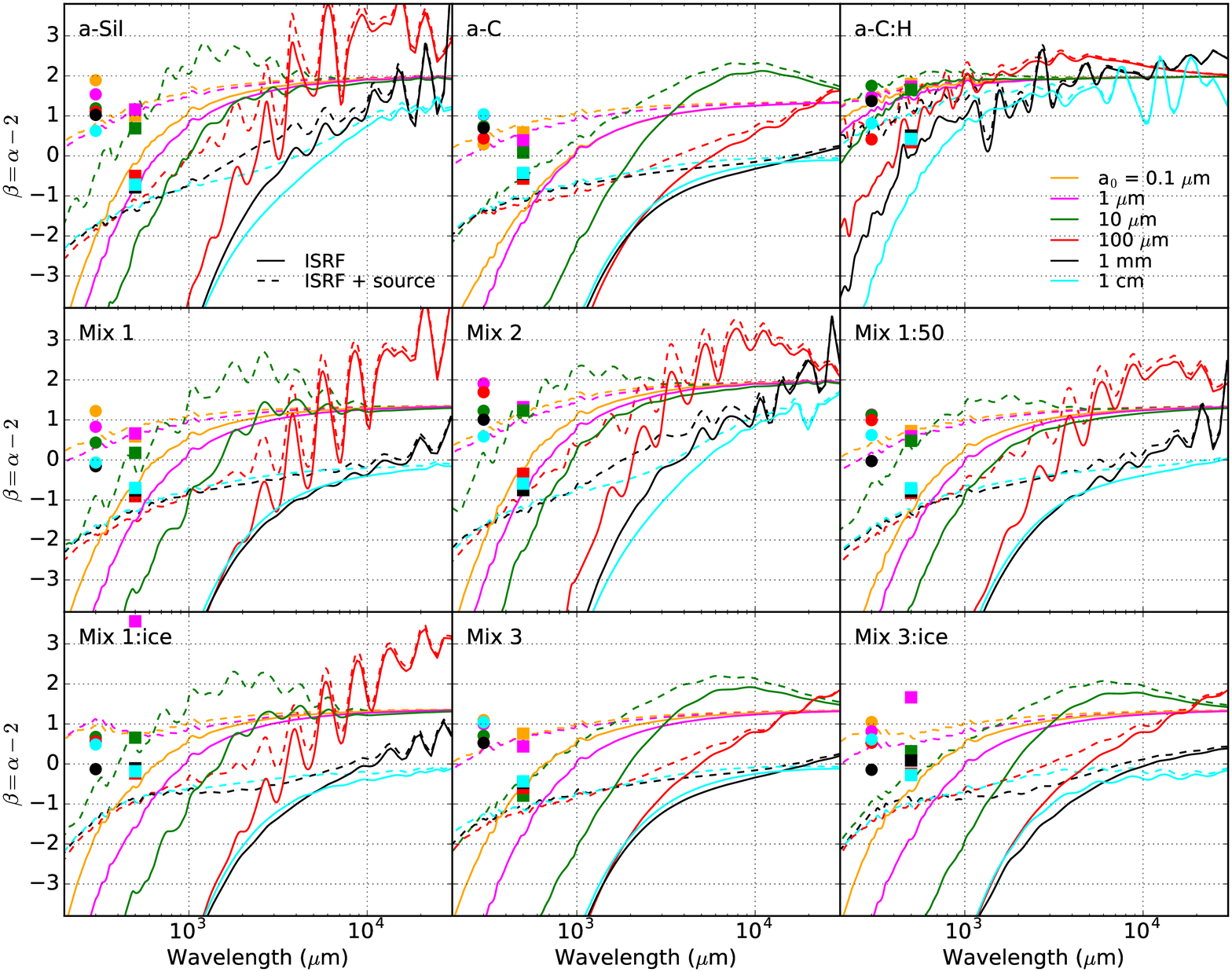}}
\caption{Spectral indices of the SEDs presented in Fig.~\ref{figure_CRT_SEDs} (optically thick medium), calculated over 1~GHz frequency steps (lines). The symbols display the spectral indices calculated using a modified blackbody in the Wien part of the SEDs. The first case, clouds illuminated by the standard ISRF alone, is displayed by solid lines and circles, and the second case, clouds illuminated by the ISRF in addition to an internal heating source, by dashed lines and squares. The same colour code is used as in Fig.~\ref{figure_CRT_SEDs}.}
\label{figure_CRT_SEDs_slopes} 
\end{figure*}

Compared to the optically thin case, the SEDs of the dense clouds illuminated by the ISRF alone, regardless of grain composition, are slightly redshifted (solid lines in Fig.~\ref{figure_CRT_SEDs}). The shift in wavelength decreases with increasing $a_0$ values. As a result, their spectral indices are somewhat lower at short wavelengths but tend asymptotically towards the same value at long wavelengths, that of the grain absorption efficiency $Q_{abs}$ (Figs.~\ref{figure_efficiencies} and \ref{figure_CRT_SEDs_slopes}). The fact that there is little difference between the optically thin and optically thick cases shows that the SEDs of the latter are dominated by the outer layers of the clouds where the grains are the warmest \citep[e.g.][and references therein]{Ysard2012}. In prestellar cores, the density distribution can be steeper than what we consider here \citep[e.g.][]{Tafalla2002, Tafalla2004}. To test this, we ran calculations with $q = 3$ and $n_C = 10^5$~H/cm$^3$ (see Eq.~\ref{equation_rho}) and found that the spectral index does not vary. The only difference is that the SED is slightly lower than in the $q = 2$ case due to a lower column density.

As expected, grains heated by an additional internal source are hotter, hence the shift of the SED peak positions towards shorter wavelengths (dashed lines in Fig.~\ref{figure_CRT_SEDs}). The SEDs are also broader than for clouds illuminated only externally by the ISRF. In this latter case, the SEDs are dominated by the outer layers of the clouds, whereas for internally heated clouds, the inner parts are hot enough to contribute significantly to the total emitted flux resulting in a broader dust temperature distribution and thus in wider SEDs. It can be seen in Fig.~\ref{figure_CRT_SEDs} that the SEDs are composed of two components: a `cold' component that corresponds to the cloud outer layers and very closely resembles the ISRF-only case, and a `hot' component responsible for the shorter-wavelength peak corresponding to the  inner layers of the heated cloud. For compositions including water ice mantles (Mix 1:ice and Mix 3:ice, see bottom left and right plots in Fig.~\ref{figure_CRT_SEDs}), the SEDs appear distorted at the peak for $a_0 = 0.1$ and 1~$\mu$m. This distortion is due to the water ice bands at $\sim 13$ and 50~$\mu$m that become visible because these small grains are relatively hot. Again, the spectral indices for all size distributions and compositions tend asymptotically towards that of their material absorption efficiencies at long wavelengths (Fig.~\ref{figure_CRT_SEDs_slopes}). Yet, since the SEDs peak at shorter wavelengths and are wider, there are two significant differences between the case with internal source compared to the case without: (i) the spectral index asymptotic value is reached at shorter wavelengths - for example $\lambda \sim 1$~mm instead of 1~cm for $a_0 \leqslant 1~\mu$m; (ii) for the largest grains, the spectral index is systematically higher with $\beta = \alpha - 2 > -1$ for $\lambda > 1$~mm. As for the case of clouds without an internal source, we tested the influence of changing the density distribution steepness $q$ by calculating the SEDs for $q = 3$ and $n_C = 10^5$~H/cm$^3$ (see Eq.~\ref{equation_rho}). Since the density decreases faster from the inside out, the grains are globally hotter. The difference in the SED peak position is negligible for $a_0 \leqslant 10~\mu$m and of the order of 100~$\mu$m for larger $a_0$. The resulting spectral indices are thus strictly similar to those of the $q = 2$ case for small $a_0$ values at all wavelengths and at $\lambda \gtrsim 1$~mm for larger $a_0$.

In addition, Fig.~\ref{figure_CRT_SEDs_slopes} also shows the spectral indices obtained by fitting the Wien part of the SEDs with a modified blackbody (see Sect.~\ref{optically_thin} for details). Compared to the optically thin case, the results differ significantly only when an internal source is considered. Due to the broader temperature distribution along the line of sight, the spectral indices are lower for all size distributions and even become negative for the largest log-normal centroids for most dust compositions. For the two compositions with water ice mantles (Mix 1:ice and Mix 3:ice), the distortion of the SED peaks due to the 13 and 50~$\mu$m ice bands produces higher spectral indices for the two smallest values of $a_0$: $\beta(a_0 = 0.1~\mu{\rm m}) = 5.6$ and $\beta(a_0 = 1~\mu{\rm m}) = 3.6$ in the Mix 1:ice case, and $\beta(a_0 = 0.1~\mu{\rm m}) = 7.0$ and $\beta(a_0 = 1~\mu{\rm m}) = 1.7$ in the Mix 3:ice case.

\subsection{Spectral indices in ALMA and VLA bands}
\label{ALMA}

\begin{figure*}[!th]
\centerline{\begin{tabular}{cc}
\includegraphics[width=0.45\textwidth]{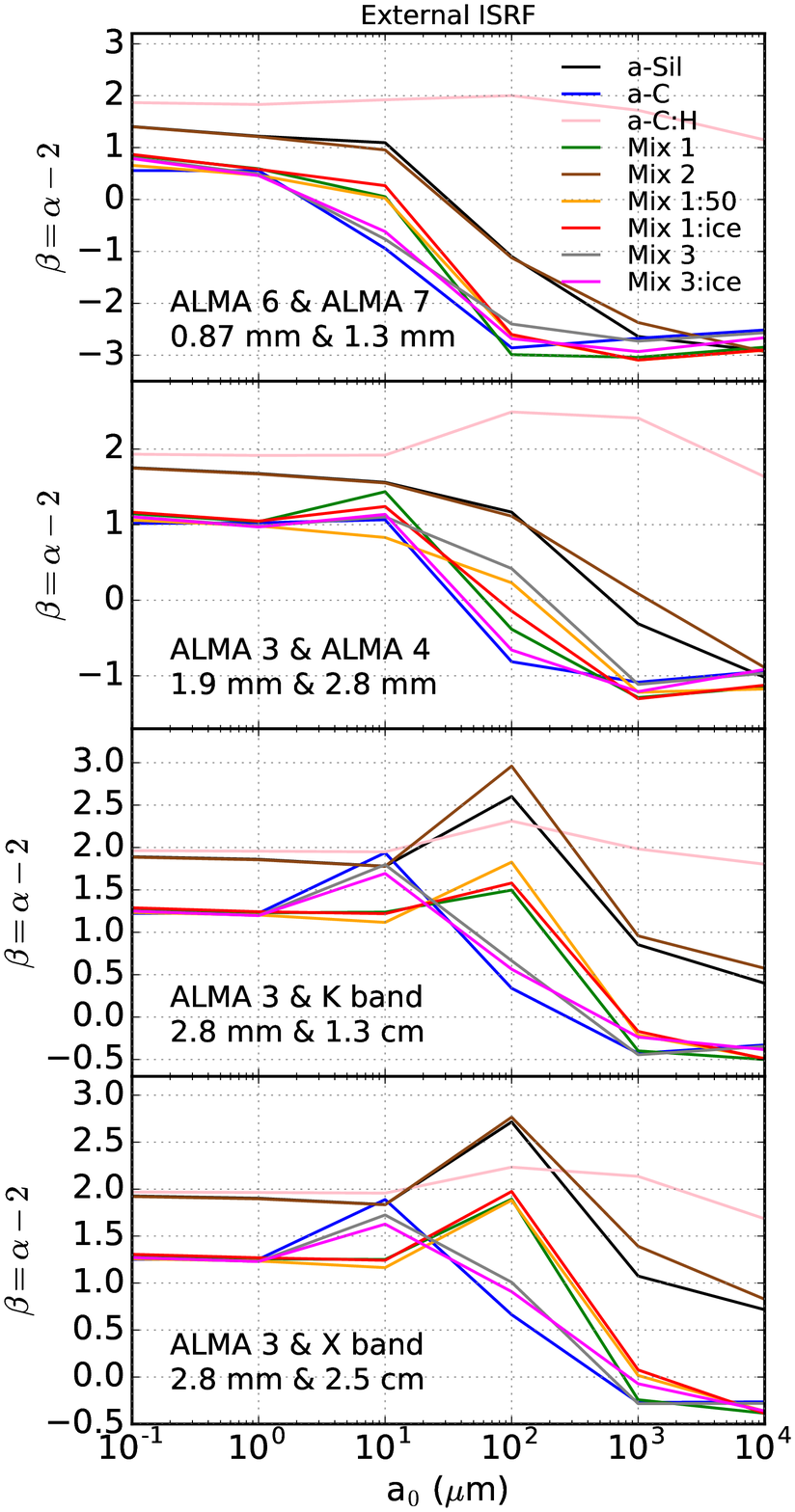} & \includegraphics[width=0.45\textwidth]{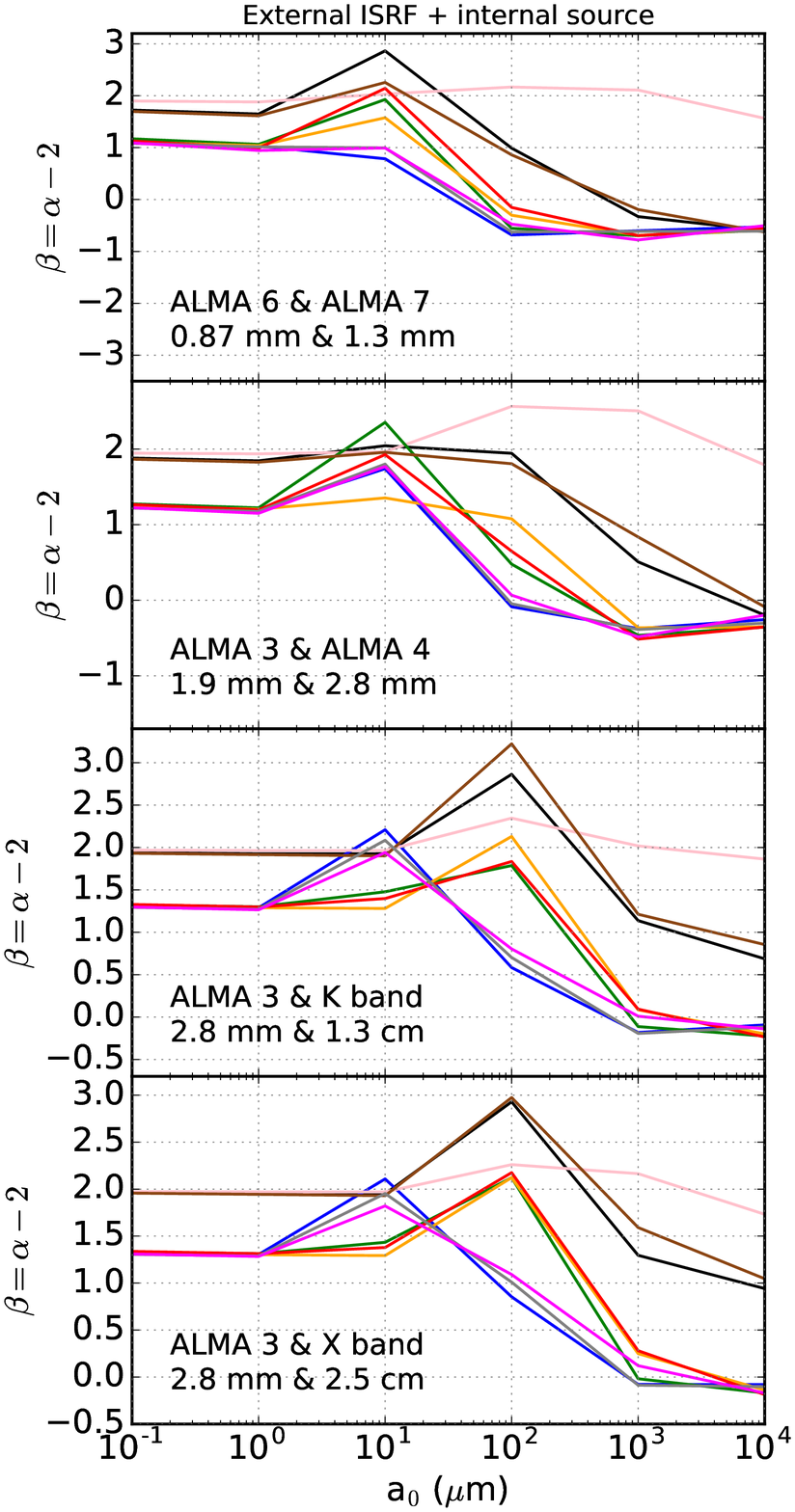}
\end{tabular}}
\caption{Spectral indices of the SEDs presented in Fig.~\ref{figure_CRT_SEDs} obtained considering a cloud with central density $n_C = 10^5$~H/cm$^3$ (Eq.~\ref{equation_rho}). From top to bottom, the spectral indices are computed between ALMA bands 6 and 7 (centred at 1.3 and 0.87 mm, respectively), ALMA bands 3 and 4 (centred at 2.8 and 1.9 mm, respectively), ALMA band 3 and VLA K band at 23~GHz ($\equiv 1.3~$cm), and ALMA band 3 and VLA X band at 12~GHz ($\equiv 2.5~$cm). Same colour code as in Fig.~\ref{figure_ISRF_temperatures} for the grain composition. {\it Left:} Cloud externally illuminated by the standard ISRF alone. {\it Right:} Cloud with an additional internal heating source (see Sect.~\ref{dust_emission} for details).}
\label{figure_slopes_ALMA} 
\end{figure*}

Now that we have a clear picture of the shape of the dust SED and of its variations in optically thick regions, we present the spectral indices calculated between ALMA and VLA bands in Fig.~\ref{figure_slopes_ALMA} for the same dust and cloud characteristics as in the previous Sect.~\ref{optically_thick}. The spectral indices are calculated between ALMA bands 3 and 4 (centred at 2.8 and 1.9~mm, respectively), between ALMA bands 6 and 7 (centred at 1.3 and 0.87~mm, respectively), between ALMA band 3 and VLA K-band (1.3~cm), and between ALMA band 3 and VLA X-band (2.5~cm).

The case of a cloud externally heated by the ISRF only is shown in the left plots in Fig.~\ref{figure_slopes_ALMA}, where the spectral indices calculated between ALMA bands 6 and 7 are always lower than those calculated between bands 3 and 4. Indeed, the former band positions roughly match the SED peak position whereas the latter ones are in the Rayleigh domain. In both cases, the spectral indices decrease with increasing size distribution centroid in agreement with Fig.~\ref{figure_CRT_SEDs_slopes}. For bands 6 and 7, $0 \leqslant \beta \leqslant 1$ implies that $a_0 \lesssim 10~\mu$m and $\beta < 0$ otherwise. For bands 3 and 4, $\beta$ is greater than one when $a_0 \lesssim 10~\mu$m, $0 \leqslant \beta \leqslant 1$ for $10 \lesssim a_0 \lesssim 100~\mu$m and negative otherwise. When we consider the spectral indices calculated between ALMA band 3 and VLA bands K and X, we see that, for grains of mixed composition and excluding those containing a-C:H, its value depends mainly on their size. An increase in $\beta$ around $a_0 = 100~\mu$m is found as already shown in Fig.~\ref{figure_CRT_SEDs_slopes} (see also Fig.~\ref{figure_ISRF_SEDs_slopes} and associated text in Sect.~\ref{optically_thin}). In both cases, $\beta > 1$ means that $a_0 < 100~\mu$m, $0 \leqslant \beta \leqslant 1$ that $100~\mu$m $< a_0 < 1~$mm, and $\beta$ is negative when $a_0 > 1~$mm.

The case of a cloud with an internal source is shown in the right plots in Fig.~\ref{figure_slopes_ALMA}. As in the previous case, the spectral indices calculated between ALMA bands 6 and 7 are inferior to those calculated between ALMA bands 3 and 4, both being always higher than in the case of clouds only externally heated. For $a_0 = 0.1$ and 1~$\mu$m, the increase is around 0.5. A decrease in the spectral index for increasing grain sizes is again found, with one notable exception: size distributions centred at $a_0 = 10~\mu$m, for which the spectral indices are higher than for smaller and larger size distribution centroids. The magnitude of this increase depends on the grain material composition and is explained by the shape of the SEDs presented in Fig.~\ref{figure_slopes_ALMA}. From these SEDs, one can see that the connection between the contributions to the total emitted flux of the cloud inner and outer layers (see Sect.~\ref{optically_thick}) arises between 1 and 3~mm, thus matching the positions of the considered ALMA bands. The exact size distribution centroid at which this effect becomes visible depends mostly on the internal source characteristics. However, despite the simplicity of our toy model, this clearly illustrates the difficulty of assessing dust grain sizes only from the millimetre slope of the SED in YSOs. In the test case shown in Fig.~\ref{figure_slopes_ALMA}, such high spectral index values could be mistaken for grain size distributions centred on smaller grains. For spectral indices calculated between ALMA band 3 and VLA bands K and X, the trends are similar to those of clouds without an internal source. The only difference is slightly higher values for $a_0 > 100~\mu$m, with $\beta \sim 0 \pm 0.2$.

\begin{figure*}[!th]
\centerline{\begin{tabular}{cc}
\includegraphics[width=0.45\textwidth]{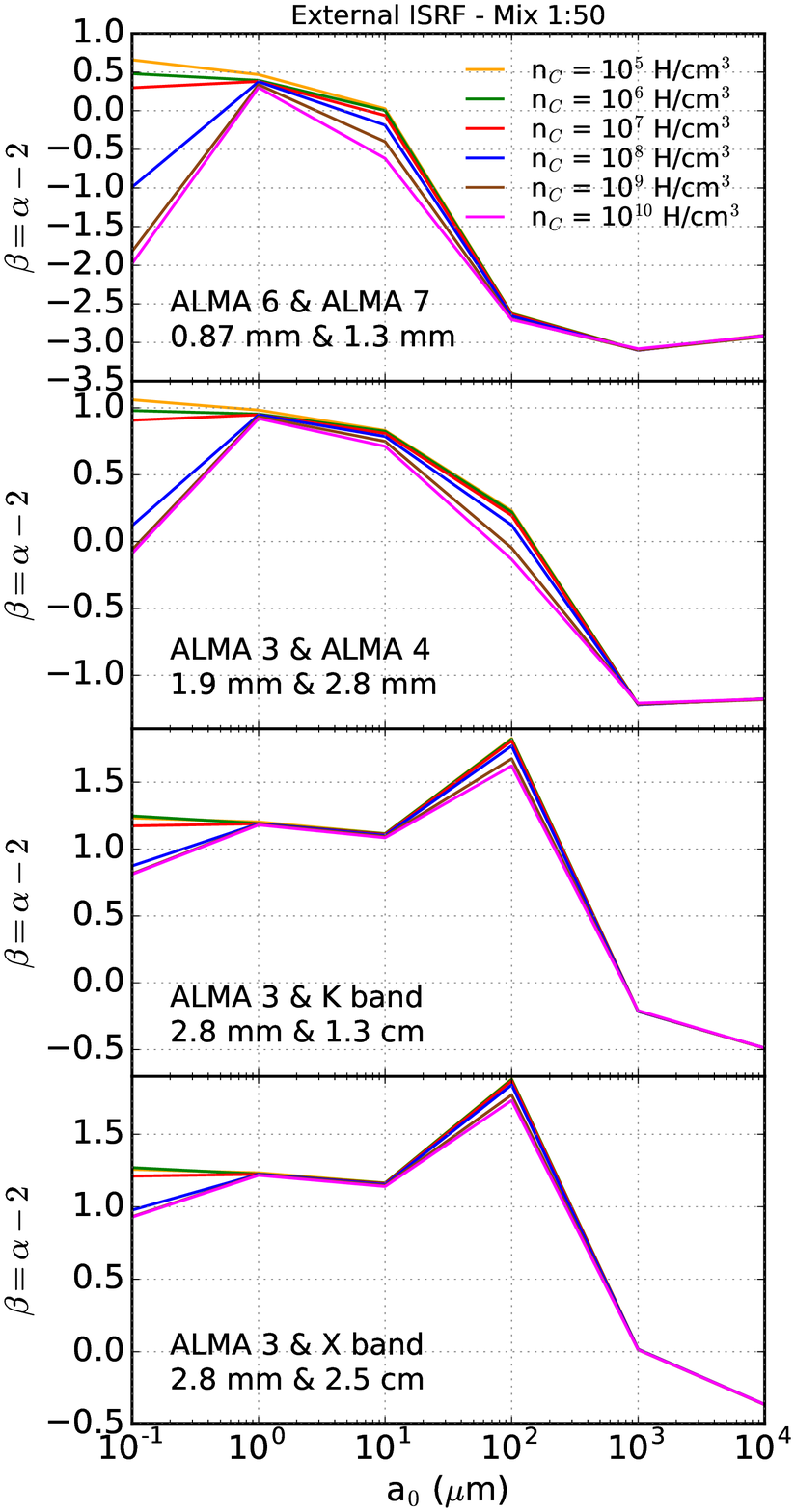} & \includegraphics[width=0.45\textwidth]{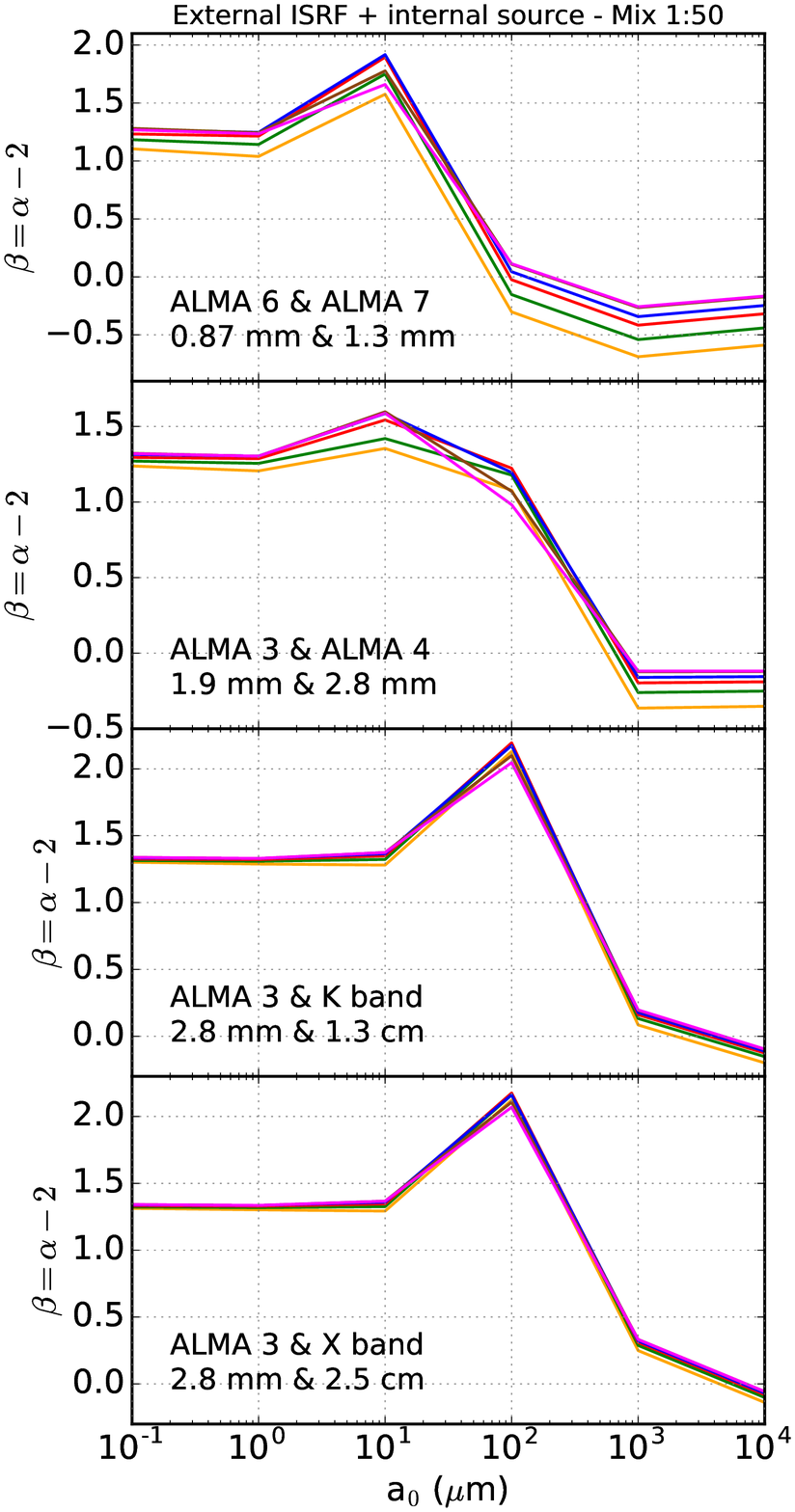}
\end{tabular}}
\caption{Spectral indices of SEDs obtained considering clouds with Mix 1:50 grains and central densities $n_C = 10^5, 10^6, 10^7, 10^8, 10^9$, and $10^{10}$~H/cm$^3$ (Eq.~\ref{equation_rho}), plotted in orange, green, red, blue, brown, and magenta, respectively. From top to bottom, the spectral indices are computed between ALMA bands 6 and 7 (centred at 1.3 and 0.87 mm, respectively), ALMA bands 3 and 4 (centred at 2.8 and 1.9 mm, respectively), ALMA band 3 and K band at 23~GHz ($\equiv 1.3~$cm), and ALMA band 3 and X band at 12~GHz ($\equiv 2.5~$cm). {\it Left:} Cloud externally illuminated by the standard ISRF alone. {\it Right:} Cloud with an additional internal heating source (see Sect.~\ref{dust_emission} for details).}
\label{figure_slopes_ALMA_nC} 
\end{figure*}

Figure~\ref{figure_slopes_ALMA_nC} shows the influence of the cloud central density on the spectral index measured between the same ALMA and VLA bands: six $n_C$ values are considered, $n_C = 10^5, 10^6, 10^7$, $10^8, 10^9$, and $10^{10}$~H/cm$^3$. In the case of clouds only externally illuminated by the ISRF, the cloud density influence is small for all centroid values except of the smallest one, $a_0 = 0.1~\mu$m. Indeed, only the smallest grains have heat capacities small enough to be sensitive to the small variations in the radiation field intensity at the cloud centre induced by the increase in $n_C$. Moreover, when bands at longer wavelengths are considered, the variations in $\beta$ decrease. In the case of clouds with internal sources, the differences are more significant in ALMA bands: from 0.12 to 0.27 for spectral indices calculated between bands 3 and 4, and from 0.06 to 0.16 between bands 6 and 7. This results from the decrease in the radiation field intensity at the cloud centres with increasing $n_C$ leading to lower grain temperatures and thus slightly shallower dust SEDs. As in the case without internal source, the variations with $n_C$ are reduced to almost nothing when one considers the spectral indices calculated between the ALMA band 3 and the VLA bands. This implies that in the case of observations for which no information on the density of the region is available, it is preferable to consider the longest possible wavelengths to determine the grain size.

\section{Conclusion}
\label{summary}

The aim of this paper is to understand the influence of the grain composition and size distribution on the shape of their SED in dense interstellar regions such as molecular clouds, prestellar cores, YSOs, and protoplanetary discs. Three main results emerge from this study.
\begin{enumerate}
\item Modified blackbody fits of dust SEDs in combination with the choice of a mass absorption coefficient reference value at a given FIR/submillimetre/millimetre wavelength is often used to determine dust masses and subsequently interstellar object masses. It is well known that the mass absorption coefficient strongly depends on the dust composition, leading to errors on mass estimates up to a factor of four for the compositions considered in our study. We emphasise that the choice of the mass absorption coefficient is also strongly dependent on the dust size distribution, potentially leading to errors greater than one order of magnitude. 
\item It appears almost impossible to determine the exact composition of dust grains from the FIR/submillimetre/millimetre SED shape alone. Indeed, from one composition to another, the variation in the position of the SED peak does not seem significant enough to be discriminatory unless one has very precise knowledge of the local radiation field and the geometry of the region. Moreover, it can be noted that for all grains containing carbon, regardless of its proportion to silicate, their spectral index is completely determined by this carbon (valid for both a-C and a-C:H). In addition, neither porosity nor the presence of ice mantles influence the steepness of the slope. Combining information at these wavelengths with observations at shorter wavelengths, and in particular with data showing spectral features, characteristics of the materials expected in dense ISM regions may lift the degeneracies (something that could be done with the instruments onboard the James Webb Space Telescope - JWST - and the SPace Infrared telescope for Cosmology and Astrophysics - SPICA). Even if not presented in this paper, the spectral shape of the water ice and silicate bands also depends on the dust grains exact composition (and shape). 
\item Lastly, as expected, determining the grain size from the SED (sub)millimetre to centimetre spectral index alone is not an easy task. In particular, using the `true' opacity spectral index value directly from the optical properties may not always be a good idea. The spectral index of the SED can indeed be strongly modified by the dust temperature distribution along the line of sight, in particular when observing regions with internal heating sources and/or complex geometries that were beyond the scope of this study (i.e. YSOs and protoplanetary discs). Finally, considering the longest possible wavelengths is better to ensure that observations are done in the Rayleigh domain (far from the SED peak) and in optically thin conditions.
\end{enumerate}
In conclusion, the determination of grain masses, sizes, and compositions from only (sub)millimetre to centimetre continuum data is highly uncertain, a situation that can be improved with complementary observations in the mid- to far-infrared.

\begin{acknowledgements}
We would like to thank A. Dutrey, A. Maury, and S. Guilloteau for fruitful discussions. We also thank M. Juvela and V. Ossenkopf for making their softwares publicly available (for radiative transfer and dust optical properties calculations, respectively). Finally, we thank our anonymous referee whose careful reading and interesting comments helped to clarify and improve the paper. I. Jimenez-Serra acknowledges the support from the STFC through an Ernest Rutherford fellowship (grant number ST/L004801), from the FEDER funding under grant ESP2017-86582-C4-1-R, and from the Spanish State Research Agency (AEI) through project number MDM-2017-0737 Unidad de Excelencia “María de Maeztu”- Centro de Astrobiología (INTA-CSIC). A.P. Jones, L. Verstraete, and N. Ysard acknowledge the support of the Programme National PCMI of CNRS/INSU with INC/INP co-funded by CEA and CNES.
\end{acknowledgements}

\bibliography{biblio}

\begin{thebibliography}{92}
\expandafter\ifx\csname natexlab\endcsname\relax\def\natexlab#1{#1}\fi

\bibitem[{{Agladze} {et~al.}(1996){Agladze}, {Sievers}, {Jones}, {Burlitch}, \&
  {Beckwith}}]{Agladze1996}
{Agladze}, N.~I., {Sievers}, A.~J., {Jones}, S.~A., {Burlitch}, J.~M., \&
  {Beckwith}, S.~V.~W. 1996, \apj, 462, 1026

\bibitem[{{Andrews} {et~al.}(2009){Andrews}, {Wilner}, {Hughes}, {Qi}, \&
  {Dullemond}}]{Andrews2009}
{Andrews}, S.~M., {Wilner}, D.~J., {Hughes}, A.~M., {Qi}, C., \& {Dullemond},
  C.~P. 2009, \apj, 700, 1502

\bibitem[{{Arzoumanian} {et~al.}(2011){Arzoumanian}, {Andr{\'e}}, {Didelon},
  {K{\"o}nyves}, {Schneider}, {Men'shchikov}, {Sousbie}, {Zavagno}, {Bontemps},
  {di Francesco}, {Griffin}, {Hennemann}, {Hill}, {Kirk}, {Martin}, {Minier},
  {Molinari}, {Motte}, {Peretto}, {Pezzuto}, {Spinoglio}, {Ward-Thompson},
  {White}, \& {Wilson}}]{Arzoumanian2011}
{Arzoumanian}, D., {Andr{\'e}}, P., {Didelon}, P., {et~al.} 2011, \aap, 529, L6

\bibitem[{{Beckwith} {et~al.}(1990){Beckwith}, {Sargent}, {Chini}, \&
  {Guesten}}]{Beckwith1990}
{Beckwith}, S.~V.~W., {Sargent}, A.~I., {Chini}, R.~S., \& {Guesten}, R. 1990,
  \aj, 99, 924

\bibitem[{{Birnstiel} {et~al.}(2018){Birnstiel}, {Dullemond}, {Zhu}, {Andrews},
  {Bai}, {Wilner}, {Carpenter}, {Huang}, {Isella}, {Benisty}, {P{\'e}rez}, \&
  {Zhang}}]{Birnstiel2018}
{Birnstiel}, T., {Dullemond}, C.~P., {Zhu}, Z., {et~al.} 2018, \apjl, 869, L45

\bibitem[{{Birnstiel} {et~al.}(2011){Birnstiel}, {Ormel}, \&
  {Dullemond}}]{Birnstiel2011}
{Birnstiel}, T., {Ormel}, C.~W., \& {Dullemond}, C.~P. 2011, \aap, 525, A11

\bibitem[{{Birnstiel} {et~al.}(2010){Birnstiel}, {Ricci}, {Trotta},
  {Dullemond}, {Natta}, {Testi}, {Dominik}, {Henning}, {Ormel}, \&
  {Zsom}}]{Birnstiel2010}
{Birnstiel}, T., {Ricci}, L., {Trotta}, F., {et~al.} 2010, \aap, 516, L14

\bibitem[{{Blum}(2018)}]{Blum2018}
{Blum}, J. 2018, \ssr, 214, 52

\bibitem[{{Blum} \& {Wurm}(2008)}]{Blum2008}
{Blum}, J. \& {Wurm}, G. 2008, \araa, 46, 21

\bibitem[{Bohren \& Huffman(1983)}]{Bohren1983}
Bohren, C. \& Huffman, D. 1983, Absorption and Scattering of Light by Small
  Particles (Wiley and Sons: New York -- Chichester -- Brisbane -- Toronto --
  Singapore)

\bibitem[{{Bohren} \& {Huffman}(1998)}]{BHMIE}
{Bohren}, C.~F. \& {Huffman}, D.~R. 1998, {Absorption and Scattering of Light
  by Small Particles}, 544

\bibitem[{{Boudet} {et~al.}(2005){Boudet}, {Mutschke}, {Nayral}, {J{\"a}ger},
  {Bernard}, {Henning}, \& {Meny}}]{Boudet2005}
{Boudet}, N., {Mutschke}, H., {Nayral}, C., {et~al.} 2005, \apj, 633, 272

\bibitem[{{Busquet} {et~al.}(2019){Busquet}, {Girart}, {Estalella},
  {Fern{\'a}ndez-L{\'o}pez}, {Galv{\'a}n-Madrid}, {Anglada},
  {Carrasco-Gonz{\'a}lez}, {A{\~n}ez-L{\'o}pez}, {Curiel}, {Osorio},
  {Rodr{\'{\i}}guez}, \& {Torrelles}}]{Busquet2019}
{Busquet}, G., {Girart}, J.~M., {Estalella}, R., {et~al.} 2019, \aap, 623, L8

\bibitem[{{Campeggio} {et~al.}(2007){Campeggio}, {Strafella}, {Maiolo}, {Elia},
  \& {Aiello}}]{Campeggio2007}
{Campeggio}, L., {Strafella}, F., {Maiolo}, B., {Elia}, D., \& {Aiello}, S.
  2007, \apj, 668, 316

\bibitem[{{Chiang} {et~al.}(2012){Chiang}, {Looney}, \& {Tobin}}]{Chiang2012}
{Chiang}, H.-F., {Looney}, L.~W., \& {Tobin}, J.~J. 2012, \apj, 756, 168

\bibitem[{{Choi} {et~al.}(2017){Choi}, {Kang}, {Lee}, {Tatematsu}, {Kang},
  {Sayers}, {Evans}, {Cho}, {Kwon}, {Park}, {Ohashi}, {Yoo}, \&
  {Lee}}]{Choi2017}
{Choi}, M., {Kang}, M., {Lee}, J.-E., {et~al.} 2017, \apjs, 232, 24

\bibitem[{{Compi{\`e}gne} {et~al.}(2011){Compi{\`e}gne}, {Verstraete}, {Jones},
  {Bernard}, {Boulanger}, {Flagey}, {Le Bourlot}, {Paradis}, \&
  {Ysard}}]{Compiegne2011}
{Compi{\`e}gne}, M., {Verstraete}, L., {Jones}, A., {et~al.} 2011, \aap, 525,
  A103

\bibitem[{{Coupeaud} {et~al.}(2011){Coupeaud}, {Demyk}, {Meny}, {Nayral},
  {Delpech}, {Leroux}, {Depecker}, {Creff}, {Brubach}, \& {Roy}}]{Coupeaud2011}
{Coupeaud}, A., {Demyk}, K., {Meny}, C., {et~al.} 2011, \aap, 535, A124

\bibitem[{{D'Alessio} {et~al.}(2001){D'Alessio}, {Calvet}, \&
  {Hartmann}}]{Alessio2001}
{D'Alessio}, P., {Calvet}, N., \& {Hartmann}, L. 2001, \apj, 553, 321

\bibitem[{{Davis} \& {Ryan}(1990)}]{Davis1990}
{Davis}, D.~R. \& {Ryan}, E.~V. 1990, \icarus, 83, 156

\bibitem[{{Demyk} {et~al.}(2017){Demyk}, {Meny}, {Lu}, {Papatheodorou},
  {Toplis}, {Leroux}, {Depecker}, {Brubach}, {Roy}, {Nayral}, {Ojo}, {Delpech},
  {Paradis}, \& {Gromov}}]{Demyk2017}
{Demyk}, K., {Meny}, C., {Lu}, X.-H., {et~al.} 2017, \aap, 600, A123

\bibitem[{{Dominik} {et~al.}(2016){Dominik}, {Paszun}, \&
  {Borel}}]{Dominik2016}
{Dominik}, C., {Paszun}, D., \& {Borel}, H. 2016, ArXiv e-prints

\bibitem[{{Draine}(2006)}]{Draine2006}
{Draine}, B.~T. 2006, \apj, 636, 1114

\bibitem[{{Gerin} {et~al.}(2017){Gerin}, {Pety}, {Commer{\c c}on}, {Fuente},
  {Cernicharo}, {Marcelino}, {Ciardi}, {Lis}, {Roueff}, {Wootten}, \&
  {Chapillon}}]{Gerin2017}
{Gerin}, M., {Pety}, J., {Commer{\c c}on}, B., {et~al.} 2017, \aap, 606, A35

\bibitem[{{Guilloteau} {et~al.}(2011){Guilloteau}, {Dutrey}, {Pi{\'e}tu}, \&
  {Boehler}}]{Guilloteau2011}
{Guilloteau}, S., {Dutrey}, A., {Pi{\'e}tu}, V., \& {Boehler}, Y. 2011, \aap,
  529, A105

\bibitem[{{Gundlach} \& {Blum}(2015)}]{Gundlach2015}
{Gundlach}, B. \& {Blum}, J. 2015, \apj, 798, 34

\bibitem[{{G{\"u}ttler} {et~al.}(2010){G{\"u}ttler}, {Blum}, {Zsom}, {Ormel},
  \& {Dullemond}}]{Guttler2010}
{G{\"u}ttler}, C., {Blum}, J., {Zsom}, A., {Ormel}, C.~W., \& {Dullemond},
  C.~P. 2010, \aap, 513, A56

\bibitem[{{G{\"u}ttler} {et~al.}(2009){G{\"u}ttler}, {Krause}, {Geretshauser},
  {Speith}, \& {Blum}}]{Guttler2009}
{G{\"u}ttler}, C., {Krause}, M., {Geretshauser}, R.~J., {Speith}, R., \&
  {Blum}, J. 2009, \apj, 701, 130

\bibitem[{{Isella} {et~al.}(2010){Isella}, {Carpenter}, \&
  {Sargent}}]{Isella2010}
{Isella}, A., {Carpenter}, J.~M., \& {Sargent}, A.~I. 2010, \apj, 714, 1746

\bibitem[{{Isella} {et~al.}(2012){Isella}, {P{\'e}rez}, \&
  {Carpenter}}]{Isella2012}
{Isella}, A., {P{\'e}rez}, L.~M., \& {Carpenter}, J.~M. 2012, \apj, 747, 136

\bibitem[{{Jones}(2012{\natexlab{a}})}]{Jones2012a}
{Jones}, A.~P. 2012{\natexlab{a}}, A\&A, 540, A1

\bibitem[{{Jones}(2012{\natexlab{b}})}]{Jones2012b}
{Jones}, A.~P. 2012{\natexlab{b}}, A\&A, 540, A2

\bibitem[{{Jones}(2012{\natexlab{c}})}]{Jones2012c}
{Jones}, A.~P. 2012{\natexlab{c}}, A\&A, 542, A98

\bibitem[{{Jones} {et~al.}(2013){Jones}, {Fanciullo}, {K{\"o}hler},
  {Verstraete}, {Guillet}, {Bocchio}, \& {Ysard}}]{Jones2013}
{Jones}, A.~P., {Fanciullo}, L., {K{\"o}hler}, M., {et~al.} 2013, \aap, 558,
  A62

\bibitem[{{Jones} {et~al.}(2017){Jones}, {K{\"o}hler}, {Ysard}, {Bocchio}, \&
  {Verstraete}}]{Jones2017}
{Jones}, A.~P., {K{\"o}hler}, M., {Ysard}, N., {Bocchio}, M., \& {Verstraete},
  L. 2017, \aap, 602, A46

\bibitem[{{Jones} {et~al.}(2016){Jones}, {K{\"o}hler}, {Ysard}, {Dartois},
  {Godard}, \& {Gavilan}}]{Jones2016}
{Jones}, A.~P., {K{\"o}hler}, M., {Ysard}, N., {et~al.} 2016, \aap, 588, A43

\bibitem[{{Juvela}(2005)}]{Juvela2005}
{Juvela}, M. 2005, \aap, 440, 531

\bibitem[{{Juvela} {et~al.}(2015){Juvela}, {Demyk}, {Doi}, {Hughes},
  {Lef{\`e}vre}, {Marshall}, {Meny}, {Montillaud}, {Pagani}, {Paradis},
  {Ristorcelli}, {Malinen}, {Montier}, {Paladini}, {Pelkonen}, \&
  {Rivera-Ingraham}}]{Juvela2015}
{Juvela}, M., {Demyk}, K., {Doi}, Y., {et~al.} 2015, \aap, 584, A94

\bibitem[{{Juvela} \& {Padoan}(2003)}]{Juvela2003}
{Juvela}, M. \& {Padoan}, P. 2003, \aap, 397, 201

\bibitem[{{Kataoka} {et~al.}(2013){Kataoka}, {Tanaka}, {Okuzumi}, \&
  {Wada}}]{Kataoka2013}
{Kataoka}, A., {Tanaka}, H., {Okuzumi}, S., \& {Wada}, K. 2013, \aap, 557, L4

\bibitem[{{K{\"o}hler} {et~al.}(2011){K{\"o}hler}, {Guillet}, \&
  {Jones}}]{Koehler2011}
{K{\"o}hler}, M., {Guillet}, V., \& {Jones}, A. 2011, \aap, 528, A96

\bibitem[{{K{\"o}hler} {et~al.}(2014){K{\"o}hler}, {Jones}, \&
  {Ysard}}]{Koehler2014}
{K{\"o}hler}, M., {Jones}, A., \& {Ysard}, N. 2014, \aap, 565, L9

\bibitem[{{K{\"o}hler} {et~al.}(2012){K{\"o}hler}, {Stepnik}, {Jones},
  {Guillet}, {Abergel}, {Ristorcelli}, \& {Bernard}}]{Koehler2012}
{K{\"o}hler}, M., {Stepnik}, B., {Jones}, A.~P., {et~al.} 2012, \aap, 548, A61

\bibitem[{{K{\"o}hler} {et~al.}(2015){K{\"o}hler}, {Ysard}, \&
  {Jones}}]{Koehler2015}
{K{\"o}hler}, M., {Ysard}, N., \& {Jones}, A.~P. 2015, \aap, 579, A15

\bibitem[{{Kwon} {et~al.}(2015){Kwon}, {Looney}, {Mundy}, \&
  {Welch}}]{Kwon2015}
{Kwon}, W., {Looney}, L.~W., {Mundy}, L.~G., \& {Welch}, W.~J. 2015, \apj, 808,
  102

\bibitem[{{Liu} {et~al.}(2015){Liu}, {Yin}, {Hu}, {Jin}, \&
  {Sorensen}}]{Liu2015}
{Liu}, C., {Yin}, Y., {Hu}, F., {Jin}, H., \& {Sorensen}, C.~M. 2015, Aerosol
  Science Technology, 49, 928

\bibitem[{{Liu} {et~al.}(2017){Liu}, {Henning}, {Carrasco-Gonz{\'a}lez},
  {Chandler}, {Linz}, {Birnstiel}, {van Boekel}, {P{\'e}rez}, {Flock}, {Testi},
  {Rodr{\'{\i}}guez}, \& {Galv{\'a}n-Madrid}}]{Liu2017}
{Liu}, Y., {Henning}, T., {Carrasco-Gonz{\'a}lez}, C., {et~al.} 2017, \aap,
  607, A74

\bibitem[{{Lorek} {et~al.}(2018){Lorek}, {Lacerda}, \& {Blum}}]{Lorek2018}
{Lorek}, S., {Lacerda}, P., \& {Blum}, J. 2018, \aap, 611, A18

\bibitem[{{Mathis} {et~al.}(1983){Mathis}, {Mezger}, \& {Panagia}}]{Mathis1983}
{Mathis}, J.~S., {Mezger}, P.~G., \& {Panagia}, N. 1983, \aap, 128, 212

\bibitem[{{Maxwell Garnett}(1904)}]{MG1904}
{Maxwell Garnett}, J. 1904, Royal Society of London Philosophical Transactions
  Series A, 203

\bibitem[{{Mennella} {et~al.}(1998){Mennella}, {Brucato}, {Colangeli},
  {Palumbo}, {Rotundi}, \& {Bussoletti}}]{Mennella1998}
{Mennella}, V., {Brucato}, J.~R., {Colangeli}, L., {et~al.} 1998, \apj, 496,
  1058

\bibitem[{{Meny} {et~al.}(2007){Meny}, {Gromov}, {Boudet}, {Bernard},
  {Paradis}, \& {Nayral}}]{Meny2007}
{Meny}, C., {Gromov}, V., {Boudet}, N., {et~al.} 2007, \aap, 468, 171

\bibitem[{{Mie}(1908)}]{Mie1908}
{Mie}, G. 1908, Annalen der Physik, 330, 377

\bibitem[{{Miettinen} {et~al.}(2012){Miettinen}, {Harju}, {Haikala}, \&
  {Juvela}}]{Miettinen2012}
{Miettinen}, O., {Harju}, J., {Haikala}, L.~K., \& {Juvela}, M. 2012, \aap,
  538, A137

\bibitem[{{Min} {et~al.}(2016){Min}, {Rab}, {Woitke}, {Dominik}, \&
  {M{\'e}nard}}]{Min2016}
{Min}, M., {Rab}, C., {Woitke}, P., {Dominik}, C., \& {M{\'e}nard}, F. 2016,
  \aap, 585, A13

\bibitem[{{Miotello} {et~al.}(2014){Miotello}, {Testi}, {Lodato}, {Ricci},
  {Rosotti}, {Brooks}, {Maury}, \& {Natta}}]{Miotello2014}
{Miotello}, A., {Testi}, L., {Lodato}, G., {et~al.} 2014, \aap, 567, A32

\bibitem[{{Mishchenko} {et~al.}(2016{\natexlab{a}}){Mishchenko}, {Dlugach}, \&
  {Liu}}]{Mishchenko2016a}
{Mishchenko}, M.~I., {Dlugach}, J.~M., \& {Liu}, L. 2016{\natexlab{a}}, \jqsrt,
  178, 284

\bibitem[{{Mishchenko} {et~al.}(2016{\natexlab{b}}){Mishchenko}, {Dlugach},
  {Yurkin}, {Bi}, {Cairns}, {Liu}, {Panetta}, {Travis}, {Yang}, \&
  {Zakharova}}]{Mishchenko2016b}
{Mishchenko}, M.~I., {Dlugach}, J.~M., {Yurkin}, M.~A., {et~al.}
  2016{\natexlab{b}}, \physrep, 632, 1

\bibitem[{{Natta} \& {Testi}(2004)}]{Natta2004}
{Natta}, A. \& {Testi}, L. 2004, in Astronomical Society of the Pacific
  Conference Series, Vol. 323, Star Formation in the Interstellar Medium: In
  Honor of David Hollenbach, ed. D.~{Johnstone}, F.~C. {Adams}, D.~N.~C. {Lin},
  D.~A. {Neufeeld}, \& E.~C. {Ostriker}, 279

\bibitem[{{Natta} {et~al.}(2007){Natta}, {Testi}, {Calvet}, {Henning},
  {Waters}, \& {Wilner}}]{Natta2007}
{Natta}, A., {Testi}, L., {Calvet}, N., {et~al.} 2007, Protostars and Planets
  V, 767

\bibitem[{{Ostriker}(1964)}]{Ostriker1964}
{Ostriker}, J. 1964, \apj, 140, 1056

\bibitem[{{Pagani} {et~al.}(2010){Pagani}, {Steinacker}, {Bacmann}, {Stutz}, \&
  {Henning}}]{Pagani2010}
{Pagani}, L., {Steinacker}, J., {Bacmann}, A., {Stutz}, A., \& {Henning}, T.
  2010, Science, 329, 1622

\bibitem[{{Paszun} \& {Dominik}(2009)}]{Paszun2009}
{Paszun}, D. \& {Dominik}, C. 2009, \aap, 507, 1023

\bibitem[{{P{\'e}rez} {et~al.}(2012){P{\'e}rez}, {Carpenter}, {Chandler},
  {Isella}, {Andrews}, {Ricci}, {Calvet}, {Corder}, {Deller}, {Dullemond},
  {Greaves}, {Harris}, {Henning}, {Kwon}, {Lazio}, {Linz}, {Mundy}, {Sargent},
  {Storm}, {Testi}, \& {Wilner}}]{Perez2012}
{P{\'e}rez}, L.~M., {Carpenter}, J.~M., {Chandler}, C.~J., {et~al.} 2012,
  \apjl, 760, L17

\bibitem[{{P{\'e}rez} {et~al.}(2015){P{\'e}rez}, {Chandler}, {Isella},
  {Carpenter}, {Andrews}, {Calvet}, {Corder}, {Deller}, {Dullemond}, {Greaves},
  {Harris}, {Henning}, {Kwon}, {Lazio}, {Linz}, {Mundy}, {Ricci}, {Sargent},
  {Storm}, {Tazzari}, {Testi}, \& {Wilner}}]{Perez2015}
{P{\'e}rez}, L.~M., {Chandler}, C.~J., {Isella}, A., {et~al.} 2015, \apj, 813,
  41

\bibitem[{{Planck Collaboration} {et~al.}(2011){Planck Collaboration}, {Ade},
  {Aghanim}, {Arnaud}, {Ashdown}, {Aumont}, {Baccigalupi}, {Balbi}, {Banday},
  {Barreiro}, \& et~al.}]{PLXXII}
{Planck Collaboration}, {Ade}, P.~A.~R., {Aghanim}, N., {et~al.} 2011, \aap,
  536, A22

\bibitem[{{Pollack} {et~al.}(1994){Pollack}, {Hollenbach}, {Beckwith},
  {Simonelli}, {Roush}, \& {Fong}}]{Pollack1994}
{Pollack}, J.~B., {Hollenbach}, D., {Beckwith}, S., {et~al.} 1994, \apj, 421,
  615

\bibitem[{{Remy} {et~al.}(2017){Remy}, {Grenier}, {Marshall}, \&
  {Casandjian}}]{Remy2017}
{Remy}, Q., {Grenier}, I.~A., {Marshall}, D.~J., \& {Casandjian}, J.~M. 2017,
  \aap, 601, A78

\bibitem[{{Remy} {et~al.}(2018){Remy}, {Grenier}, {Marshall}, \&
  {Casandjian}}]{Remy2018}
{Remy}, Q., {Grenier}, I.~A., {Marshall}, D.~J., \& {Casandjian}, J.~M. 2018,
  ArXiv e-prints

\bibitem[{{Ribas} {et~al.}(2017){Ribas}, {Espaillat}, {Mac{\'{\i}}as}, {Bouy},
  {Andrews}, {Calvet}, {Naylor}, {Riviere-Marichalar}, {van der Wiel}, \&
  {Wilner}}]{Ribas2017}
{Ribas}, {\'A}., {Espaillat}, C.~C., {Mac{\'{\i}}as}, E., {et~al.} 2017, \apj,
  849, 63

\bibitem[{{Schlafly} {et~al.}(2016){Schlafly}, {Meisner}, {Stutz},
  {Kainulainen}, {Peek}, {Tchernyshyov}, {Rix}, {Finkbeiner}, {Covey}, {Green},
  {Bell}, {Burgett}, {Chambers}, {Draper}, {Flewelling}, {Hodapp}, {Kaiser},
  {Magnier}, {Martin}, {Metcalfe}, {Wainscoat}, \& {Waters}}]{Schlafly2016}
{Schlafly}, E.~F., {Meisner}, A.~M., {Stutz}, A.~M., {et~al.} 2016, \apj, 821,
  78

\bibitem[{{Scott} \& {Duley}(1996)}]{Scott1996}
{Scott}, A. \& {Duley}, W.~W. 1996, \apjs, 105, 401

\bibitem[{{Sheehan} \& {Eisner}(2017)}]{Sheehan2017}
{Sheehan}, P.~D. \& {Eisner}, J.~A. 2017, \apj, 851, 45

\bibitem[{{Steinacker} {et~al.}(2015){Steinacker}, {Andersen}, {Thi},
  {Paladini}, {Juvela}, {Bacmann}, {Pelkonen}, {Pagani}, {Lef{\`e}vre},
  {Henning}, \& {Noriega-Crespo}}]{Steinacker2015}
{Steinacker}, J., {Andersen}, M., {Thi}, W.-F., {et~al.} 2015, \aap, 582, A70

\bibitem[{{Tafalla} {et~al.}(2004){Tafalla}, {Myers}, {Caselli}, \&
  {Walmsley}}]{Tafalla2004}
{Tafalla}, M., {Myers}, P.~C., {Caselli}, P., \& {Walmsley}, C.~M. 2004, \aap,
  416, 191

\bibitem[{{Tafalla} {et~al.}(2002){Tafalla}, {Myers}, {Caselli}, {Walmsley}, \&
  {Comito}}]{Tafalla2002}
{Tafalla}, M., {Myers}, P.~C., {Caselli}, P., {Walmsley}, C.~M., \& {Comito},
  C. 2002, \apj, 569, 815

\bibitem[{{Tanaka} {et~al.}(1996){Tanaka}, {Inaba}, \& {Nakazawa}}]{Tanaka1996}
{Tanaka}, H., {Inaba}, S., \& {Nakazawa}, K. 1996, \icarus, 123, 450

\bibitem[{{Tazaki} \& {Tanaka}(2018)}]{Tazaki2018}
{Tazaki}, R. \& {Tanaka}, H. 2018, ArXiv e-prints

\bibitem[{{Tazzari} {et~al.}(2016){Tazzari}, {Testi}, {Ercolano}, {Natta},
  {Isella}, {Chandler}, {P{\'e}rez}, {Andrews}, {Wilner}, {Ricci}, {Henning},
  {Linz}, {Kwon}, {Corder}, {Dullemond}, {Carpenter}, {Sargent}, {Mundy},
  {Storm}, {Calvet}, {Greaves}, {Lazio}, \& {Deller}}]{Tazzari2016}
{Tazzari}, M., {Testi}, L., {Ercolano}, B., {et~al.} 2016, \aap, 588, A53

\bibitem[{{Tripathi} {et~al.}(2018){Tripathi}, {Andrews}, {Birnstiel},
  {Chandler}, {Isella}, {P{\'e}rez}, {Harris}, {Ricci}, {Wilner}, {Carpenter},
  {Calvet}, {Corder}, {Deller}, {Dullemond}, {Greaves}, {Henning}, {Kwon},
  {Lazio}, {Linz}, \& {Testi}}]{Tripathi2018}
{Tripathi}, A., {Andrews}, S.~M., {Birnstiel}, T., {et~al.} 2018, \apj, 861, 64

\bibitem[{{Trotta} {et~al.}(2013){Trotta}, {Testi}, {Natta}, {Isella}, \&
  {Ricci}}]{Trotta2013}
{Trotta}, F., {Testi}, L., {Natta}, A., {Isella}, A., \& {Ricci}, L. 2013,
  \aap, 558, A64

\bibitem[{{Wada} {et~al.}(2008){Wada}, {Tanaka}, {Suyama}, {Kimura}, \&
  {Yamamoto}}]{Wada2008}
{Wada}, K., {Tanaka}, H., {Suyama}, T., {Kimura}, H., \& {Yamamoto}, T. 2008,
  \apj, 677, 1296

\bibitem[{{Warren}(1984)}]{Warren1984}
{Warren}, S.~G. 1984, AO, 23, 1206

\bibitem[{{Weidling} {et~al.}(2009){Weidling}, {G{\"u}ttler}, {Blum}, \&
  {Brauer}}]{Weidling2009}
{Weidling}, R., {G{\"u}ttler}, C., {Blum}, J., \& {Brauer}, F. 2009, \apj, 696,
  2036

\bibitem[{{Whittet} {et~al.}(2001){Whittet}, {Gerakines}, {Hough}, \&
  {Shenoy}}]{Whittet2001}
{Whittet}, D.~C.~B., {Gerakines}, P.~A., {Hough}, J.~H., \& {Shenoy}, S.~S.
  2001, \apj, 547, 872

\bibitem[{{Windmark} {et~al.}(2012){Windmark}, {Birnstiel}, {G{\"u}ttler},
  {Blum}, {Dullemond}, \& {Henning}}]{Windmark2012}
{Windmark}, F., {Birnstiel}, T., {G{\"u}ttler}, C., {et~al.} 2012, \aap, 540,
  A73

\bibitem[{{Wu} {et~al.}(2016){Wu}, {Cheng}, {Zheng}, \& {Chen}}]{Wu2016}
{Wu}, Y., {Cheng}, T., {Zheng}, L., \& {Chen}, H. 2016, \jqsrt, 168, 158

\bibitem[{{Wuchterl} \& {Tscharnuter}(2003)}]{Wuchterl2003}
{Wuchterl}, G. \& {Tscharnuter}, W.~M. 2003, \aap, 398, 1081

\bibitem[{{Young} {et~al.}(2018){Young}, {Bate}, {Mowat}, {Hatchell}, \&
  {Harries}}]{Young2018}
{Young}, A.~K., {Bate}, M.~R., {Mowat}, C.~F., {Hatchell}, J., \& {Harries},
  T.~J. 2018, \mnras, 474, 800

\bibitem[{{Ysard} {et~al.}(2018){Ysard}, {Jones}, {Demyk}, {Bout{\'e}raon}, \&
  {Koehler}}]{Ysard2018}
{Ysard}, N., {Jones}, A.~P., {Demyk}, K., {Bout{\'e}raon}, T., \& {Koehler}, M.
  2018, \aap, 617, A124

\bibitem[{{Ysard} {et~al.}(2012){Ysard}, {Juvela}, {Demyk}, {Guillet},
  {Abergel}, {Bernard}, {Malinen}, {M{\'e}ny}, {Montier}, {Paradis},
  {Ristorcelli}, \& {Verstraete}}]{Ysard2012}
{Ysard}, N., {Juvela}, M., {Demyk}, K., {et~al.} 2012, \aap, 542, A21

\bibitem[{{Ysard} {et~al.}(2016){Ysard}, {K{\"o}hler}, {Jones}, {Dartois},
  {Godard}, \& {Gavilan}}]{Ysard2016}
{Ysard}, N., {K{\"o}hler}, M., {Jones}, A., {et~al.} 2016, \aap, 588, A44

\end{thebibliography}

\begin{appendix}

\section{Mass absorption coefficient}
\label{appendix_kappa}

Tables~\ref{table_kappa1} to \ref{table_kappa9} shows the mass absorption coefficient $\kappa$ values for the dust compositions presented in Table~\ref{table_composition} in the case of log-normal and power-law size distributions. The $\kappa$ values are presented for six wavelengths: 250 and 500~$\mu$m, 0.87, 1.3, 1.9, and 2.8~mm. The trends are similar to those described in Sect.~\ref{section_kappa} for all wavelengths and compositions, even if it should be noticed that, in the case of a log-normal size distribution, the peak size of $\kappa$ depends on the considered wavelength.

\begin{table*}[ht]
\centering
\caption{Mass absorption coefficient $\kappa$ for grains made of a-Sil at $\lambda = 250$ and 500~$\mu$m, and 0.87, 1.3, 1.9, and 2.8~mm. The values are given in the cases of: (i) a log-normal size distribution with $a_0 = 0.1, 1, 10, 50, 100,$ and 500~$\mu$m, 1 and 5~mm, and 1~cm for $0.01~\mu$m $\leqslant a \leqslant 10$~cm ; (ii) a power-law size distribution with $p = -4, -3.5, -3, -2.5$, and -2 for $0.01~\mu$m $\leqslant a \leqslant 10$~cm ; (iii) a power-law size distribution with $p = -3.5$ for $0.01~\mu$m $\leqslant a \leqslant a_{max}$ where $a_{max} = 0.1, 1, 10$ and 100~$\mu$m, 1~mm, and 1~cm.} 
\label{table_kappa1}
\begin{tabular}{c|cccccc}
\hline
\multicolumn{7}{c}{Dust composition: a-Sil} \\
\hline
Size distribution & \multicolumn{6}{c}{Mass absorption coefficient $\kappa_{\lambda}$ (cm$^2$/g)} \\
\hline
Log-normal $a_0$  & $\kappa_{250~\mu{\rm m}}$ & $\kappa_{500~\mu{\rm m}}$ & $\kappa_{0.87~{\rm mm}}$ & $\kappa_{1.3~{\rm mm}}$ & $\kappa_{1.9~{\rm mm}}$ & $\kappa_{2.8~{\rm mm}}$ \\
$0.1~\mu$m & 5.33  & 1.33  & 0.43 & 0.19 & 0.09 & 0.04 \\
$1~\mu$m   & 5.39  & 1.33  & 0.43 & 0.19 & 0.09 & 0.04 \\
$10~\mu$m  & 21.34 & 2.37  & 0.51 & 0.20 & 0.10 & 0.04 \\
$50~\mu$m  & 41.01 & 12.48 & 3.19 & 0.85 & 0.26 & 0.07 \\
$100~\mu$m & 23.59 & 12.22 & 4.88 & 1.82 & 0.69 & 0.17 \\
$500~\mu$m & 3.78  & 3.69  & 2.73 & 1.91 & 1.14 & 0.57 \\
$1~$mm     & 1.72  & 1.80  & 1.57 & 1.26 & 0.85 & 0.53 \\
$5~$mm     & 0.30  & 0.31  & 0.32 & 0.31 & 0.28 & 0.22 \\
$1~$cm     & 0.15  & 0.15  & 0.16 & 0.16 & 0.15 & 0.13 \\
\hline
Power-law $p$  & $\kappa_{250~\mu{\rm m}}$ & $\kappa_{500~\mu{\rm m}}$ & $\kappa_{0.87~{\rm mm}}$ & $\kappa_{1.3~{\rm mm}}$ & $\kappa_{1.9~{\rm mm}}$ & $\kappa_{2.8~{\rm mm}}$ \\
-4   & 5.33 & 1.33 & 0.43 & 0.19 & 0.09 & 0.04 \\
-3.5 & 5.33 & 1.33 & 0.43 & 0.19 & 0.09 & 0.04 \\
-3   & 5.35 & 1.33 & 0.43 & 0.19 & 0.09 & 0.04 \\
-2.5 & 5.95 & 1.46 & 0.47 & 0.21 & 0.10 & 0.04 \\
-2   & 9.50 & 2.87 & 1.13 & 0.56 & 0.30 & 0.15 \\
\hline
Power-law $a_{max}$  & $\kappa_{250~\mu{\rm m}}$ & $\kappa_{500~\mu{\rm m}}$ & $\kappa_{0.87~{\rm mm}}$ & $\kappa_{1.3~{\rm mm}}$ & $\kappa_{1.9~{\rm mm}}$ & $\kappa_{2.8~{\rm mm}}$ \\
$0.1~\mu$m & 8.24 & 2.14 & 0.73 & 0.33 & 0.16 & 0.07 \\
$1~\mu$m   & 6.30 & 1.55 & 0.50 & 0.22 & 0.11 & 0.05 \\
$10~\mu$m  & 5.57 & 1.37 & 0.45 & 0.20 & 0.10 & 0.04 \\
$100~\mu$m & 5.38 & 1.34 & 0.44 & 0.19 & 0.09 & 0.04 \\
$100~\mu$m & 5.34 & 1.33 & 0.43 & 0.19 & 0.09 & 0.04 \\
$1~$mm     & 5.33 & 1.33 & 0.43 & 0.19 & 0.09 & 0.04 \\
$1~$cm     & 5.33 & 1.33 & 0.43 & 0.19 & 0.09 & 0.04 \\
\end{tabular}
\end{table*}

\begin{table*}[ht]
\centering
\caption{Same as Table~\ref{table_kappa1} but for grains made of a-C.} 
\label{table_kappa2}
\begin{tabular}{c|cccccc}
\hline
\multicolumn{7}{c}{Dust composition: a-C} \\
\hline
Size distribution & \multicolumn{6}{c}{Mass absorption coefficient $\kappa_{\lambda}$ (cm$^2$/g)} \\
\hline
Log-normal $a_0$  & $\kappa_{250~\mu{\rm m}}$ & $\kappa_{500~\mu{\rm m}}$ & $\kappa_{0.87~{\rm mm}}$ & $\kappa_{1.3~{\rm mm}}$ & $\kappa_{1.9~{\rm mm}}$ & $\kappa_{2.8~{\rm mm}}$ \\
$0.1~\mu$m & 12.21  & 4.83  & 2.25  & 1.30  & 0.80  & 0.46 \\
$1~\mu$m   & 15.00  & 5.22  & 2.33  & 1.32  & 0.81  & 0.46 \\
$10~\mu$m  & 128.80 & 44.77 & 14.48 & 5.70  & 2.45  & 0.96 \\
$50~\mu$m  & 63.47  & 52.41 & 36.69 & 24.95 & 16.19 & 8.79 \\
$100~\mu$m & 28.76  & 28.81 & 25.42 & 21.03 & 16.51 & 11.47 \\
$500~\mu$m & 4.15   & 4.27  & 4.47  & 4.62  & 4.67  & 4.53 \\
$1~$mm     & 5.33   & 1.33  & 0.43  & 0.19  & 0.09  & 0.04 \\
$5~$mm     & 0.34   & 0.33  & 0.32  & 0.31  & 0.31  & 0.31 \\
$1~$cm     & 0.17   & 0.16  & 0.15  & 0.15  & 0.15  & 0.14 \\
\hline
Power-law $p$  & $\kappa_{250~\mu{\rm m}}$ & $\kappa_{500~\mu{\rm m}}$ & $\kappa_{0.87~{\rm mm}}$ & $\kappa_{1.3~{\rm mm}}$ & $\kappa_{1.9~{\rm mm}}$ & $\kappa_{2.8~{\rm mm}}$ \\
-4   & 12.19 & 4.83  & 2.25 & 1.29 & 0.80 & 0.46 \\
-3.5 & 12.20 & 4.83  & 2.25 & 1.30 & 0.80 & 0.46 \\
-3   & 12.35 & 4.87  & 2.27 & 1.30 & 0.80 & 0.46 \\
-2.5 & 15.36 & 6.07  & 2.82 & 1.61 & 0.99 & 0.59\\
-2   & 26.60 & 13.09 & 7.29 & 4.78 & 3.30 & 2.16 \\
\hline
Power-law $a_{max}$  & $\kappa_{250~\mu{\rm m}}$ & $\kappa_{500~\mu{\rm m}}$ & $\kappa_{0.87~{\rm mm}}$ & $\kappa_{1.3~{\rm mm}}$ & $\kappa_{1.9~{\rm mm}}$ & $\kappa_{2.8~{\rm mm}}$ \\
$0.1~\mu$m & 23.86 & 10.32 & 5.16 & 3.12 & 2.01 & 1.22 \\
$1~\mu$m   & 16.82 & 6.72  & 3.14 & 1.81 & 1.12 & 0.65 \\
$10~\mu$m  & 13.60 & 5.33  & 2.46 & 1.41 & 0.87 & 0.50 \\
$100~\mu$m & 12.58 & 4.95  & 2.30 & 1.32 & 0.81 & 0.47 \\
$100~\mu$m & 12.29 & 4.86  & 2.26 & 1.30 & 0.80 & 0.46 \\
$1~$mm     & 12.22 & 4.83  & 2.25 & 1.30 & 0.80 & 0.46 \\
$1~$cm     & 12.20 & 4.83  & 2.25 & 1.30 & 0.80 & 0.46 \\
\end{tabular}
\end{table*}

\begin{table*}[ht]
\centering
\caption{Same as Table~\ref{table_kappa1} but for grains made of a-C:H.} 
\label{table_kappa3}
\begin{tabular}{c|cccccc}
\hline
\multicolumn{7}{c}{Dust composition: a-C:H} \\
\hline
Size distribution & \multicolumn{6}{c}{Mass absorption coefficient $\kappa_{\lambda}$ (cm$^2$/g)} \\
\hline
Log-normal $a_0$  & $\kappa_{250~\mu{\rm m}}$ & $\kappa_{500~\mu{\rm m}}$ & $\kappa_{0.87~{\rm mm}}$ & $\kappa_{1.3~{\rm mm}}$ & $\kappa_{1.9~{\rm mm}}$ & $\kappa_{2.8~{\rm mm}}$ \\
$0.1~\mu$m & 5.81$\times 10^{-4}$ & 1.50$\times 10^{-4}$ & 4.88$\times 10^{-5}$ & 2.17$\times 10^{-5}$ & 1.06$\times 10^{-5}$ & 4.73$\times 10^{-6}$ \\
$1~\mu$m   & 5.83$\times 10^{-4}$ & 1.50$\times 10^{-4}$ & 4.88$\times 10^{-5}$ & 2.17$\times 10^{-5}$ & 1.06$\times 10^{-5}$ & 4.73$\times 10^{-6}$ \\
$10~\mu$m  & 7.57$\times 10^{-4}$ & 1.63$\times 10^{-4}$ & 5.02$\times 10^{-5}$ & 2.19$\times 10^{-5}$ & 1.07$\times 10^{-5}$ & 4.75$\times 10^{-6}$ \\
$50~\mu$m  & 1.79$\times 10^{-3}$ & 3.20$\times 10^{-4}$ & 7.56$\times 10^{-5}$ & 2.77$\times 10^{-5}$ & 1.22$\times 10^{-5}$ & 5.05$\times 10^{-6}$ \\
$100~\mu$m & 2.13$\times 10^{-3}$ & 4.56$\times 10^{-4}$ & 1.12$\times 10^{-4}$ & 3.92$\times 10^{-5}$ & 1.57$\times 10^{-5}$ & 5.87$\times 10^{-6}$ \\
$500~\mu$m & 1.72$\times 10^{-3}$ & 5.40$\times 10^{-4}$ & 1.86$\times 10^{-4}$ & 1.03$\times 10^{-4}$ & 3.84$\times 10^{-5}$ & 1.31$\times 10^{-5}$ \\
$1~$mm     & 1.53$\times 10^{-3}$ & 4.52$\times 10^{-4}$ & 1.72$\times 10^{-4}$ & 1.59$\times 10^{-4}$ & 5.52$\times 10^{-5}$ & 1.58$\times 10^{-5}$ \\
$5~$mm     & 1.41$\times 10^{-3}$ & 3.70$\times 10^{-4}$ & 1.30$\times 10^{-4}$ & 8.55$\times 10^{-5}$ & 4.19$\times 10^{-5}$ & 1.45$\times 10^{-5}$ \\
$1$cm      & 1.37$\times 10^{-3}$ & 3.64$\times 10^{-4}$ & 1.22$\times 10^{-4}$ & 5.83$\times 10^{-5}$ & 3.13$\times 10^{-5}$ & 1.37$\times 10^{-5}$ \\
\hline
Power-law $p$  & $\kappa_{250~\mu{\rm m}}$ & $\kappa_{500~\mu{\rm m}}$ & $\kappa_{0.87~{\rm mm}}$ & $\kappa_{1.3~{\rm mm}}$ & $\kappa_{1.9~{\rm mm}}$ & $\kappa_{2.8~{\rm mm}}$ \\
-4   & 5.81$\times 10^{-4}$ & 1.50$\times 10^{-4}$ & 4.88$\times 10^{-5}$ & 2.17$\times 10^{-5}$ & 1.06$\times 10^{-5}$ & 4.73$\times 10^{-6}$ \\
-3.5 & 5.81$\times 10^{-4}$ & 1.50$\times 10^{-4}$ & 4.88$\times 10^{-5}$ & 2.17$\times 10^{-5}$ & 1.06$\times 10^{-5}$ & 4.73$\times 10^{-6}$ \\
-3   & 5.81$\times 10^{-4}$ & 1.50$\times 10^{-4}$ & 4.88$\times 10^{-5}$ & 2.17$\times 10^{-5}$ & 1.06$\times 10^{-5}$ & 4.73$\times 10^{-6}$ \\
-2.5 & 6.02$\times 10^{-4}$ & 1.54$\times 10^{-4}$ & 4.97$\times 10^{-5}$ & 2.22$\times 10^{-5}$ & 1.08$\times 10^{-5}$ & 4.78$\times 10^{-6}$ \\
-2   & 1.10$\times 10^{-3}$ & 2.78$\times 10^{-4}$ & 8.87$\times 10^{-5}$ & 4.91$\times 10^{-5}$ & 2.14$\times 10^{-5}$ & 7.99$\times 10^{-6}$ \\
\hline
Power-law $a_{max}$  & $\kappa_{250~\mu{\rm m}}$ & $\kappa_{500~\mu{\rm m}}$ & $\kappa_{0.87~{\rm mm}}$ & $\kappa_{1.3~{\rm mm}}$ & $\kappa_{1.9~{\rm mm}}$ & $\kappa_{2.8~{\rm mm}}$ \\
$0.1~\mu$m & 7.54$\times 10^{-4}$ & 1.88$\times 10^{-4}$ & 5.96$\times 10^{-5}$ & 2.86$\times 10^{-5}$ & 1.32$\times 10^{-5}$ & 5.46$\times 10^{-6}$ \\
$1~\mu$m   & 6.18$\times 10^{-4}$ & 1.57$\times 10^{-4}$ & 5.06$\times 10^{-5}$ & 2.27$\times 10^{-5}$ & 1.10$\times 10^{-5}$ & 4.83$\times 10^{-6}$ \\
$10~\mu$m  & 5.88$\times 10^{-4}$ & 1.51$\times 10^{-4}$ & 4.90$\times 10^{-5}$ & 2.18$\times 10^{-5}$ & 1.07$\times 10^{-5}$ & 4.75$\times 10^{-6}$ \\
$100~\mu$m & 5.82$\times 10^{-4}$ & 1.50$\times 10^{-4}$ & 4.88$\times 10^{-5}$ & 2.17$\times 10^{-5}$ & 1.06$\times 10^{-5}$ & 4.74$\times 10^{-6}$ \\
$100~\mu$m & 5.81$\times 10^{-4}$ & 1.50$\times 10^{-4}$ & 4.88$\times 10^{-5}$ & 2.17$\times 10^{-5}$ & 1.06$\times 10^{-5}$ & 4.73$\times 10^{-6}$ \\
$1~$mm     & 5.81$\times 10^{-4}$ & 1.50$\times 10^{-4}$ & 4.88$\times 10^{-5}$ & 2.17$\times 10^{-5}$ & 1.06$\times 10^{-5}$ & 4.73$\times 10^{-6}$ \\
$1~$cm     & 5.81$\times 10^{-4}$ & 1.50$\times 10^{-4}$ & 4.88$\times 10^{-5}$ & 2.17$\times 10^{-5}$ & 1.06$\times 10^{-5}$ & 4.73$\times 10^{-6}$ \\
\end{tabular}
\end{table*}

\begin{table*}[ht]
\centering
\caption{Same as Table~\ref{table_kappa1} but for grains made of Mix 1.} 
\label{table_kappa4}
\begin{tabular}{c|cccccc}
\hline
\multicolumn{7}{c}{Dust composition: Mix 1} \\
\hline
Size distribution & \multicolumn{6}{c}{Mass absorption coefficient $\kappa_{\lambda}$ (cm$^2$/g)} \\
\hline
Log-normal $a_0$  & $\kappa_{250~\mu{\rm m}}$ & $\kappa_{500~\mu{\rm m}}$ & $\kappa_{0.87~{\rm mm}}$ & $\kappa_{1.3~{\rm mm}}$ & $\kappa_{1.9~{\rm mm}}$ & $\kappa_{2.8~{\rm mm}}$ \\
$0.1~\mu$m & 8.82 & 3.15 & 1.39 & 0.78 & 0.47 & 0.27 \\
$1~\mu$m   & 9.07 & 3.17 & 1.40 & 0.78 & 0.47 & 0.27 \\
$10~\mu$m  & 54.26 & 10.75 & 2.33 & 0.96 & 0.51 & 0.28 \\
$50~\mu$m  & 57.51 & 41.65 & 19.37 & 9.10 & 3.33 & 1.03 \\
$100~\mu$m & 27.76 & 28.12 & 21.05 & 14.48 & 7.47 & 3.17 \\
$500~\mu$m & 3.90 & 4.36 & 4.80 & 5.01 & 4.92 & 4.39 \\
$1~$mm     & 1.77 & 1.91 & 2.09 & 2.22 & 2.34 & 2.35 \\
$5~\mu$mm  & 0.31 & 0.32 & 0.33 & 0.34 & 0.36 & 0.38 \\
$1~$cm     & 0.15 & 0.15 & 0.16 & 0.16 & 0.16 & 0.17 \\
\hline
Power-law $p$  & $\kappa_{250~\mu{\rm m}}$ & $\kappa_{500~\mu{\rm m}}$ & $\kappa_{0.87~{\rm mm}}$ & $\kappa_{1.3~{\rm mm}}$ & $\kappa_{1.9~{\rm mm}}$ & $\kappa_{2.8~{\rm mm}}$ \\
-4   & 8.82 & 3.15 & 1.39 & 0.78 & 0.47 & 0.27 \\
-3.5 & 8.82 & 3.15 & 1.39 & 0.78 & 0.47 & 0.27 \\
-3   & 8.87 & 3.16 & 1.40 & 0.78 & 0.47 & 0.27 \\
-2.5 & 10.12 & 3.64 & 1.59 & 0.89 & 0.53 & 0.30 \\
-2   & 16.10 & 7.68 & 3.97 & 2.57 & 1.62 & 1.02 \\
\hline
Power-law $a_{max}$  & $\kappa_{250~\mu{\rm m}}$ & $\kappa_{500~\mu{\rm m}}$ & $\kappa_{0.87~{\rm mm}}$ & $\kappa_{1.3~{\rm mm}}$ & $\kappa_{1.9~{\rm mm}}$ & $\kappa_{2.8~{\rm mm}}$ \\
$0.1~\mu$m & 14.35 & 5.79 & 2.67 & 1.57 & 0.93 & 0.54 \\
$1~\mu$m   & 10.80 & 3.94 & 1.73 & 0.97 & 0.57 & 0.32 \\
$10~\mu$m  & 9.35 & 3.33 & 1.46 & 0.82 & 0.49 & 0.28 \\
$100~\mu$m & 8.95 & 3.18 & 1.40 & 0.79 & 0.48 & 0.27 \\
$100~\mu$m & 8.85 & 3.15 & 1.39 & 0.78 & 0.47 & 0.27 \\
$1~$mm     & 8.83 & 3.15 & 1.39 & 0.78 & 0.47 & 0.27 \\
$1~$cm     & 8.82 & 3.15 & 1.39 & 0.78 & 0.47 & 0.27 \\
\end{tabular}
\end{table*}

\begin{table*}[ht]
\centering
\caption{Same as Table~\ref{table_kappa1} but for grains made of Mix 2.} 
\label{table_kappa5}
\begin{tabular}{c|cccccc}
\hline
\multicolumn{7}{c}{Dust composition: Mix 2} \\
\hline
Size distribution & \multicolumn{6}{c}{Mass absorption coefficient $\kappa_{\lambda}$ (cm$^2$/g)} \\
\hline
Log-normal $a_0$  & $\kappa_{250~\mu{\rm m}}$ & $\kappa_{500~\mu{\rm m}}$ & $\kappa_{0.87~{\rm mm}}$ & $\kappa_{1.3~{\rm mm}}$ & $\kappa_{1.9~{\rm mm}}$ & $\kappa_{2.8~{\rm mm}}$ \\
$0.1~\mu$m & 6.50 & 1.62 & 0.53 & 0.23 & 0.12 & 0.05 \\
$1~\mu$m   & 6.56 & 1.62 & 0.53 & 0.23 & 0.12 & 0.05 \\
$10~\mu$m  & 16.31 & 2.30 & 0.58 & 0.24 & 0.12 & 0.05 \\
$50~\mu$m  & 37.77 & 11.19 & 2.74 & 0.68 & 0.23 & 0.07 \\
$100~\mu$m & 25.70 & 12.19 & 4.77 & 1.46 & 0.59 & 0.14 \\
$500~\mu$m & 4.86 & 4.23 & 3.01 & 1.70 & 1.27 & 0.48 \\
$1~$mm     & 2.24 & 2.21 & 1.77 & 1.17 & 0.93 & 0.42 \\
$5~\mu$mm  & 0.39 & 0.41 & 0.41 & 0.37 & 0.31 & 0.20 \\
$1~$cm     & 0.19 & 0.20 & 0.20 & 0.20 & 0.18 & 0.14 \\
\hline
Power-law $p$  & $\kappa_{250~\mu{\rm m}}$ & $\kappa_{500~\mu{\rm m}}$ & $\kappa_{0.87~{\rm mm}}$ & $\kappa_{1.3~{\rm mm}}$ & $\kappa_{1.9~{\rm mm}}$ & $\kappa_{2.8~{\rm mm}}$ \\
-4   & 6.50 & 1.62 & 0.53 & 0.23 & 0.12 & 0.05 \\
-3.5 & 6.50 & 1.62 & 0.53 & 0.23 & 0.12 & 0.05 \\
-3   & 6.52 & 1.62 & 0.53 & 0.23 & 0.12 & 0.05 \\
-2.5 & 6.96 & 1.73 & 0.56 & 0.25 & 0.12 & 0.05 \\
-2   & 9.34 & 2.96 & 1.20 & 0.53 & 0.32 & 0.13 \\
\hline
Power-law $a_{max}$  & $\kappa_{250~\mu{\rm m}}$ & $\kappa_{500~\mu{\rm m}}$ & $\kappa_{0.87~{\rm mm}}$ & $\kappa_{1.3~{\rm mm}}$ & $\kappa_{1.9~{\rm mm}}$ & $\kappa_{2.8~{\rm mm}}$ \\
$0.1~\mu$m & 8.68 & 2.33 & 0.81 & 0.34 & 0.18 & 0.07 \\
$1~\mu$m   & 7.22 & 1.81 & 0.59 & 0.26 & 0.13 & 0.05 \\
$10~\mu$m  & 6.68 & 1.66 & 0.54 & 0.24 & 0.12 & 0.05 \\
$100~\mu$m & 6.54 & 1.63 & 0.53 & 0.23 & 0.12 & 0.05 \\
$100~\mu$m & 6.51 & 1.62 & 0.53 & 0.23 & 0.12 & 0.05 \\
$1~$mm     & 6.50 & 1.62 & 0.53 & 0.23 & 0.12 & 0.05 \\
$1~$cm     & 6.50 & 1.62 & 0.53 & 0.23 & 0.12 & 0.05 \\
\end{tabular}
\end{table*}

\begin{table*}[ht]
\centering
\caption{Same as Table~\ref{table_kappa1} but for grains made of Mix 1:50.} 
\label{table_kappa6}
\begin{tabular}{c|cccccc}
\hline
\multicolumn{7}{c}{Dust composition: Mix 1:50} \\
\hline
Size distribution & \multicolumn{6}{c}{Mass absorption coefficient $\kappa_{\lambda}$ (cm$^2$/g)} \\
\hline
Log-normal $a_0$  & $\kappa_{250~\mu{\rm m}}$ & $\kappa_{500~\mu{\rm m}}$ & $\kappa_{0.87~{\rm mm}}$ & $\kappa_{1.3~{\rm mm}}$ & $\kappa_{1.9~{\rm mm}}$ & $\kappa_{2.8~{\rm mm}}$ \\
$0.1~\mu$m & 28.84 & 10.47 & 4.69 & 2.65 & 1.62 & 0.92 \\
$1~\mu$m   & 29.10 & 10.49 & 4.70 & 2.65 & 1.62 & 0.92 \\
$10~\mu$m  & 64.50 & 14.92 & 5.25 & 2.77 & 1.65 & 0.93 \\
$50~\mu$m  & 103.19 & 54.12 & 19.99 & 8.08 & 3.32 & 1.33 \\
$100~\mu$m & 61.25 & 50.27 & 28.98 & 15.74 & 7.29 & 2.84 \\
$500~\mu$m & 9.50 & 10.61 & 11.15 & 10.95 & 9.34 & 7.32 \\
$1~$mm     & 4.26 & 4.66 & 5.04 & 5.27 & 5.18 & 4.89 \\
$5~$mm     & 0.74 & 0.77 & 0.80 & 0.83 & 0.87 & 0.90 \\
$1~$cm     & 0.37 & 0.37 & 0.38 & 0.39 & 0.40 & 0.42 \\
\hline
Power-law $p$  & $\kappa_{250~\mu{\rm m}}$ & $\kappa_{500~\mu{\rm m}}$ & $\kappa_{0.87~{\rm mm}}$ & $\kappa_{1.3~{\rm mm}}$ & $\kappa_{1.9~{\rm mm}}$ & $\kappa_{2.8~{\rm mm}}$ \\
-4   & 28.83 & 10.47 & 4.69 & 2.65 & 1.62 & 0.92 \\
-3.5 & 28.83 & 10.47 & 4.69 & 2.65 & 1.62 & 0.92 \\
-3   & 28.88 & 10.48 & 4.70 & 2.65 & 1.62 & 0.92 \\
-2.5 & 30.10 & 10.94 & 4.89 & 2.76 & 1.67 & 0.95 \\
-2   & 31.31 & 13.91 & 7.08 & 4.42 & 2.81 & 1.76 \\
\hline
Power-law $a_{max}$  & $\kappa_{250~\mu{\rm m}}$ & $\kappa_{500~\mu{\rm m}}$ & $\kappa_{0.87~{\rm mm}}$ & $\kappa_{1.3~{\rm mm}}$ & $\kappa_{1.9~{\rm mm}}$ & $\kappa_{2.8~{\rm mm}}$ \\
$0.1~\mu$m & 33.65 & 13.00 & 5.99 & 3.46 & 2.10 & 1.22 \\
$1~\mu$m   & 30.74 & 11.24 & 5.02 & 2.84 & 1.72 & 0.98 \\
$10~\mu$m  & 29.35 & 10.64 & 4.76 & 2.68 & 1.63 & 0.93 \\
$100~\mu$m & 28.95 & 10.50 & 4.70 & 2.66 & 1.62 & 0.92 \\
$100~\mu$m & 28.86 & 10.47 & 4.69 & 2.65 & 1.62 & 0.92 \\
$1~$mm     & 28.84 & 10.47 & 4.69 & 2.65 & 1.62 & 0.92 \\
$1~$cm     & 28.83 & 10.47 & 4.69 & 2.65 & 1.62 & 0.92 \\
\end{tabular}
\end{table*}

\begin{table*}[ht]
\centering
\caption{Same as Table~\ref{table_kappa1} but for grains made of Mix 1:ice.} 
\label{table_kappa7}
\begin{tabular}{c|cccccc}
\hline
\multicolumn{7}{c}{Dust composition: Mix 1:ice} \\
\hline
Size distribution & \multicolumn{6}{c}{Mass absorption coefficient $\kappa_{\lambda}$ (cm$^2$/g)} \\
\hline
Log-normal $a_0$  & $\kappa_{250~\mu{\rm m}}$ & $\kappa_{500~\mu{\rm m}}$ & $\kappa_{0.87~{\rm mm}}$ & $\kappa_{1.3~{\rm mm}}$ & $\kappa_{1.9~{\rm mm}}$ & $\kappa_{2.8~{\rm mm}}$ \\
$0.1~\mu$m & 14.05 & 4.84 & 2.12 & 1.18 & 0.71 & 0.40 \\
$1~\mu$m   & 14.38 & 4.87 & 2.13 & 1.18 & 0.71 & 0.40 \\
$10~\mu$m  & 71.68 & 13.64 & 3.24 & 1.39 & 0.76 & 0.41 \\
$50~\mu$m  & 82.99 & 51.18 & 22.60 & 10.08 & 4.10 & 1.26 \\
$100~\mu$m & 46.23 & 39.03 & 25.94 & 16.36 & 8.99 & 3.63 \\
$500~\mu$m & 7.23 & 8.03 & 8.43 & 8.09 & 7.37 & 5.85 \\
$1~$mm     & 3.33 & 3.54 & 3.85 & 4.02 & 4.06 & 3.72 \\
$5~\mu$mm  & 0.63 & 0.62 & 0.62 & 0.64 & 0.65 & 0.69 \\
$1~$cm     & 0.31 & 0.31 & 0.31 & 0.31 & 0.31 & 0.32 \\
\hline
Power-law $p$  & $\kappa_{250~\mu{\rm m}}$ & $\kappa_{500~\mu{\rm m}}$ & $\kappa_{0.87~{\rm mm}}$ & $\kappa_{1.3~{\rm mm}}$ & $\kappa_{1.9~{\rm mm}}$ & $\kappa_{2.8~{\rm mm}}$ \\
-4   & 14.04 & 4.84 & 2.12 & 1.18 & 0.71 & 0.40 \\
-3.5 & 14.05 & 4.84 & 2.12 & 1.18 & 0.71 & 0.40 \\
-3   & 14.11 & 4.85 & 2.13 & 1.18 & 0.71 & 0.40 \\
-2.5 & 15.76 & 5.44 & 2.37 & 1.31 & 0.78 & 0.44 \\
-2   & 23.47 & 10.51 & 5.38 & 3.38 & 2.23 & 1.37 \\
\hline
Power-law $a_{max}$  & $\kappa_{250~\mu{\rm m}}$ & $\kappa_{500~\mu{\rm m}}$ & $\kappa_{0.87~{\rm mm}}$ & $\kappa_{1.3~{\rm mm}}$ & $\kappa_{1.9~{\rm mm}}$ & $\kappa_{2.8~{\rm mm}}$ \\
$0.1~\mu$m & 21.36 & 8.13 & 3.71 & 2.13 & 1.31 & 0.74 \\
$1~\mu$m   & 16.65 & 5.81 & 2.53 & 1.40 & 0.84 & 0.47 \\
$10~\mu$m  & 14.74 & 5.06 & 2.20 & 1.22 & 0.73 & 0.41 \\
$100~\mu$m & 14.21 & 4.88 & 2.14 & 1.19 & 0.72 & 0.40 \\
$100~\mu$m & 14.08 & 4.85 & 2.13 & 1.18 & 0.71 & 0.40 \\
$1~$mm     & 14.05 & 4.84 & 2.12 & 1.18 & 0.71 & 0.40 \\
$1~$cm     & 14.05 & 4.84 & 2.12 & 1.18 & 0.71 & 0.40 \\
\end{tabular}
\end{table*}

\begin{table*}[ht]
\centering
\caption{Same as Table~\ref{table_kappa1} but for grains made of Mix 3.} 
\label{table_kappa8}
\begin{tabular}{c|cccccc}
\hline
\multicolumn{7}{c}{Dust composition: Mix 3} \\
\hline
Size distribution & \multicolumn{6}{c}{Mass absorption coefficient $\kappa_{\lambda}$ (cm$^2$/g)} \\
\hline
Log-normal $a_0$  & $\kappa_{250~\mu{\rm m}}$ & $\kappa_{500~\mu{\rm m}}$ & $\kappa_{0.87~{\rm mm}}$ & $\kappa_{1.3~{\rm mm}}$ & $\kappa_{1.9~{\rm mm}}$ & $\kappa_{2.8~{\rm mm}}$ \\
$0.1~\mu$m & 12.47 & 5.06 & 2.40 & 1.40 & 0.87 & 0.51 \\
$1~\mu$m   & 14.19 & 5.29 & 2.45 & 1.42 & 0.88 & 0.51 \\
$10~\mu$m  & 108.25 & 34.67 & 10.62 & 4.18 & 1.86 & 0.80 \\
$50~\mu$m  & 59.89 & 49.09 & 33.17 & 21.57 & 13.32 & 6.79 \\
$100~\mu$m & 27.14 & 27.48 & 24.20 & 19.71 & 15.07 & 10.01 \\
$500~\mu$m & 3.88 & 4.05 & 4.27 & 4.44 & 4.52 & 4.40 \\
$1~$mm     & 1.77 & 1.79 & 1.84 & 1.90 & 1.97 & 2.03 \\
$5~\mu$mm  & 0.31 & 0.30 & 0.30 & 0.29 & 0.29 & 0.29 \\
$1~$cm     & 0.15 & 0.15 & 0.14 & 0.14 & 0.14 & 0.14 \\
\hline
Power-law $p$  & $\kappa_{250~\mu{\rm m}}$ & $\kappa_{500~\mu{\rm m}}$ & $\kappa_{0.87~{\rm mm}}$ & $\kappa_{1.3~{\rm mm}}$ & $\kappa_{1.9~{\rm mm}}$ & $\kappa_{2.8~{\rm mm}}$ \\
-4   & 12.46 & 5.06 & 2.40 & 1.40 & 0.87 & 0.51 \\
-3.5 & 12.46 & 5.06 & 2.40 & 1.40 & 0.87 & 0.51 \\
-3   & 12.58 & 5.09 & 2.41 & 1.40 & 0.87 & 0.51 \\
-2.5 & 14.97 & 6.04 & 2.85 & 1.65 & 1.02 & 0.59 \\
-2   & 23.76 & 11.72 & 6.53 & 4.27 & 2.95 & 1.94 \\
\hline
Power-law $a_{max}$  & $\kappa_{250~\mu{\rm m}}$ & $\kappa_{500~\mu{\rm m}}$ & $\kappa_{0.87~{\rm mm}}$ & $\kappa_{1.3~{\rm mm}}$ & $\kappa_{1.9~{\rm mm}}$ & $\kappa_{2.8~{\rm mm}}$ \\
$0.1~\mu$m & 21.86 & 9.52 & 4.78 & 2.90 & 1.87 & 1.13 \\
$1~\mu$m   & 16.15 & 6.57 & 3.12 & 1.81 & 1.13 & 0.66 \\
$10~\mu$m  & 13.56 & 5.45 & 2.57 & 1.49 & 0.92 & 0.53 \\
$100~\mu$m & 12.75 & 5.15 & 2.44 & 1.42 & 0.88 & 0.51 \\
$100~\mu$m & 12.53 & 5.08 & 2.41 & 1.40 & 0.87 & 0.51 \\
$1~$mm     & 12.48 & 5.06 & 2.40 & 1.40 & 0.87 & 0.51 \\
$1~$cm     & 12.46 & 5.06 & 2.40 & 1.40 & 0.87 & 0.51 \\
\end{tabular}
\end{table*}

\begin{table*}[ht]
\centering
\caption{Same as Table~\ref{table_kappa1} but for grains made of Mix 3:ice.} 
\label{table_kappa9}
\begin{tabular}{c|cccccc}
\hline
\multicolumn{7}{c}{Dust composition: Mix 3:ice} \\
\hline
Size distribution & \multicolumn{6}{c}{Mass absorption coefficient $\kappa_{\lambda}$ (cm$^2$/g)} \\
\hline
Log-normal $a_0$  & $\kappa_{250~\mu{\rm m}}$ & $\kappa_{500~\mu{\rm m}}$ & $\kappa_{0.87~{\rm mm}}$ & $\kappa_{1.3~{\rm mm}}$ & $\kappa_{1.9~{\rm mm}}$ & $\kappa_{2.8~{\rm mm}}$ \\
$0.1~\mu$m & 19.83 & 7.84 & 3.70 & 2.15 & 1.33 & 0.77 \\
$1~\mu$m   & 21.72 & 8.09 & 3.75 & 2.16 & 1.34 & 0.78 \\
$10~\mu$m  & 136.85 & 41.80 & 12.73 & 5.13 & 2.38 & 1.08 \\
$50~\mu$m  & 89.19 & 65.45 & 41.56 & 25.94 & 15.50 & 7.67 \\
$100~\mu$m & 48.14 & 40.93 & 32.73 & 25.32 & 18.56 & 11.76 \\
$500~\mu$m & 8.04 & 8.52 & 8.51 & 8.01 & 7.36 & 6.49 \\
$1~$mm     & 3.71 & 3.70 & 3.95 & 4.04 & 3.94 & 3.65 \\
$5~\mu$mm  & 0.76 & 0.67 & 0.62 & 0.61 & 0.61 & 0.63 \\
$1~$cm     & 0.39 & 0.35 & 0.32 & 0.30 & 0.28 & 0.28 \\
\hline
Power-law $p$  & $\kappa_{250~\mu{\rm m}}$ & $\kappa_{500~\mu{\rm m}}$ & $\kappa_{0.87~{\rm mm}}$ & $\kappa_{1.3~{\rm mm}}$ & $\kappa_{1.9~{\rm mm}}$ & $\kappa_{2.8~{\rm mm}}$ \\
-4   & 19.81 & 7.83 & 3.70 & 2.15 & 1.33 & 0.77 \\
-3.5 & 19.82 & 7.83 & 3.70 & 2.15 & 1.33 & 0.77 \\
-3   & 19.96 & 7.87 & 3.71 & 2.15 & 1.34 & 0.78 \\
-2.5 & 22.94 & 9.03 & 4.24 & 2.45 & 1.52 & 0.88 \\
-2   & 33.84 & 16.24 & 8.92 & 5.79 & 3.96 & 2.57 \\
\hline
Power-law $a_{max}$  & $\kappa_{250~\mu{\rm m}}$ & $\kappa_{500~\mu{\rm m}}$ & $\kappa_{0.87~{\rm mm}}$ & $\kappa_{1.3~{\rm mm}}$ & $\kappa_{1.9~{\rm mm}}$ & $\kappa_{2.8~{\rm mm}}$ \\
$0.1~\mu$m & 31.70 & 13.42 & 6.65 & 4.00 & 2.56 & 1.53 \\
$1~\mu$m   & 24.43 & 9.69 & 4.57 & 2.65 & 1.64 & 0.95 \\
$10~\mu$m  & 21.17 & 8.31 & 3.90 & 2.25 & 1.40 & 0.81 \\
$100~\mu$m & 20.17 & 7.94 & 3.74 & 2.17 & 1.35 & 0.78 \\
$100~\mu$m & 19.91 & 7.86 & 3.71 & 2.15 & 1.34 & 0.77 \\
$1~$mm     & 19.84 & 7.84 & 3.70 & 2.15 & 1.33 & 0.77 \\
$1~$cm     & 19.82 & 7.83 & 3.70 & 2.15 & 1.33 & 0.77 \\
\end{tabular}
\end{table*}

\end{appendix}

\end{document}